\documentclass[aps,preprint,showpacs,floatfix]{revtex4}
\usepackage{graphicx,bm,amsmath,dcolumn}
\RequirePackage{color}

\definecolor{MyDarkGreen}{rgb}{0.02,0.60,0.06}

\begin{document}

	\title{Exact finite-size corrections in the dimer model on a planar square lattice.}
	
	\author{Nikolay Sh.\ Izmailian}
	\email{izmail@yerphi.am}
	\affiliation{Yerevan Physics Institute, Alikhanian Brothers 2, 375036 Yerevan, Armenia}
	\affiliation{Bogoliubov Laboratory of Theoretical Physics, Joint Institute for Nuclear Research, 141980 Dubna, Russian Federation}
		
	\author{Vladimir V. Papoyan}
	\email{vpap@theor.jinr.ru}
	\affiliation{Dubna State University, Dubna, Russian Federation}
	\affiliation{Bogoliubov Laboratory of Theoretical Physics, Joint Institute for Nuclear Research, 141980 Dubna, Russian Federation}

	\author{Robert M. Ziff}
	\email{rziff@umich.edu}
	\affiliation{Center for the Study of Complex Systems and Department of Chemical Engineering, University of Michigan,
		Ann Arbor, Michigan 48109-2136, USA}
	\date{\today}
	
	\begin{abstract}
	We consider the dimer model on the rectangular $2M \times 2N$ lattice with free boundary conditions. We derive exact expressions for the coefficients in the asymptotic expansion of the free energy in terms of the elliptic theta functions ($\theta_2, \theta_3, \theta_4$) and the elliptic integral of second kind ($E$), up to 22nd order. Surprisingly we find that ratio of the coefficients $f_p$ in the free energy expansion for strip  ($f_p^\mathrm{strip}$)  and square ($f_p^\mathrm{sq}$)  geometries $r_p={f_p^\mathrm{strip}}/{f_p^\mathrm{sq}}$ in the limit of large $p$ tends to $1/2$. Furthermore, we predict that the ratio of the coefficients $f_p$ in the free energy expansion for rectangular ($f_p(\rho)$ for aspect ratio $\rho > 1$) to the coefficients of the free energy for square geometries, multiplied by  $\rho^{-p-1}$, that is $r_p=\rho^{-p-1}  {f_p(\rho)}/{f_p^\mathrm{sq}}$,  is also equal to $1/2$ in the limit of $p \to \infty$. We find that the corner contribution to the free energy for the dimer model on rectangular $2M \times 2N$ lattices with free boundary conditions is equal to zero and explain that result in the framework of conformal field theory, in which the central charge of the considering model is $c=-2$.  We also derive a simple exact expression for the free energy of open strips of arbitrary width.
	\end{abstract}
	
	
	\maketitle

\section{Introduction}

In 1961, Kasteleyn \cite{Kasteleyn} and independently Temperley and Fisher \cite{FisherTemp} found an explicit and elegant formula for the number of dimer tilings (or perfect matchings) on an open square lattice of dimensions $2 M \times 2 N$:
\begin{equation}
Z = \prod_{j=1}^{M} \prod_{k=1}^{N} 4 \left[\cos^2\left(\frac{j \pi}{2M+1}\right) + \cos^2\left(\frac{k \pi}{2N+1}\right)\right]
\label{Kasteleyn}
\end{equation}
The asymptotic behavior of the logarithm of $N$ was also found as 
\begin{equation}
\ln Z \sim (2 M+1)(2N+1) \frac{G}{\pi} - (2 M+2 N+2) \ln(1 + \sqrt{2}) + \ln 2 + \ldots
\label{asymptotic}
\end{equation}
as $n \to \infty$, where $G$ is the Catalan constant.  
This result represents one of the fundamental exact results of statistical mechanics.  The number $Z$ is just the partition function for this system, and its logarithm gives the free energy per site, $f = \ln Z/[(2 N+1)(2M+1)]$.

The motivation of the present paper was to find higher-order contributions to this asymptotic series. With multiple precision computer algorithms such as exist in Mathematica, it is not difficult to calculate $Z$ explicitly for $N$ and $M$  in the thousands, and by looking at differences it is
possible to determine  higher-order asymptotic coefficients numerically.  Using results from \cite{Izmailian2002, Izmailian2003} and some new procedures including the automation of calculation, we were able to calculate analytically 22 terms in the asymptotic expansion of $f$].  At the same time, we found an intriguing and unexplained simple relation between the coefficients in the asymptotic series for a square, rectangle, and strip, in the limit of high order.  This paper is an account of that work.  We begin by discussing general finite-size scaling, and then go through calculations for the dimers on rectangle and strip geometries.  comparison 

\section{Finite-size scaling theory}

 Finite-size scaling theory, introduced by Fisher and Barber \cite{Fisher1972} almost five decades ago, has been of interest to scientists working on a variety of critical systems \cite{Privman1990}, and finds extensive applications in the analysis of experimental, Monte Carlo, and transfer-matrix data, as well as in theoretical developments related to conformal invariance \cite{Privman1984,Privman1988,Cardy1988,Blote1986,Affleck}. Theories of finite-size effects have been successful in deriving critical and noncritical properties of infinite systems from their finite counterparts. Exact solutions have played a key role in determining the form of the finite-size scaling. To fully understand the finite-size behaviour, it is valuable to study model systems, especially those which have exact results. Very few models of statistical mechanics have been solved exactly. The dimer model is the one of most prominent examples.

The dimer model is a classical statistical mechanics model, first  introduced to represent physical adsorption of diatomic molecules on crystal surfaces \cite{Fowler}. The  exact partition functions for the dimer model on the square $2M \times 2N$ lattice with both free and toroidal boundary conditions were obtained by Kasteleyn \cite{Kasteleyn}, Fisher \cite{Fisher}, and Temperley and Fisher \cite{FisherTemp} in 1961.

Ivashkevich, Izmailian, and Hu \cite{Izmailian2002} proposed a systematic method to compute the finite-size corrections for the partition function of free models on a torus, including the dimer model on a $2M \times 2N$ rectangular lattice, and the Ising and Gaussian models  on a $M \times N$ rectangular lattice. In particular they derive all terms of the exact asymptotic expansion of the logarithm of the partition function on a torus for a class of free exactly solvable models of
statistical mechanics. Their approach is based on an intimate relation between the terms of the asymptotic expansion and Kronecker's double series \cite{Izmailian2002}. The work \cite{Izmailian2002} has been further extended by Izmailian, Oganesyan, and Hu \cite{Izmailian2003} to the dimer model on a rectangular $M \times N$ lattice with various boundary conditions and for different parities of lattice sites in vertical $M$ and horizontal $N$ directions. Later Izmailian et al.\ \cite{Izmailian2014} derived the exact asymptotic expansions of the partition functions of the dimer model on the rectangular $(2M-1) \times (2N-1)$ lattice with a single monomer residing on the boundary under free and cylindrical boundary conditions. In particular they show that because of certain non-local features present in the model, the finite-size corrections in a crucial way depend on the parity of the lattice sites in horizontal $N$ and vertical $M$ directions. Furthermore, the change of parity of $M$ or $N$ induces a change of boundary condition \cite{Izmailian2005,Izmailian2007}.

It has been shown  \cite{Izmailian2003,Izmailian2014} that the exact asymptotic expansion of the free energy for dimers on an open rectangular $M \times N$ lattice takes the form 
\begin{equation}
f=f_{\text{bulk}}+\frac{2f_{1s}}{M}+\frac{2f_{2s}}{N}+f_{\text{corn}}\frac{\ln S}{S} +\frac{f_{0}}{S}+ \sum_{p=1}^{\infty}\frac{f_{p}}{S^{p+1}}\label{freeenergy}.
\end{equation}
Here $S=M \times N$ is the area of the lattice. The bulk free energy is $f_{\text{bulk}}$, the surface free energies are $f_{1s}$ and $f_{2s}$ and the corner free energy is $f_{\text{corn}}$.
The leading finite-size correction term is $f_0$ and the subleading correction terms are $f_p$ for $p = 1,2,3,...$. Indeed such form of the asymptotic expansion for the free energy also holds for the Ising model, the spanning tree model, the Gaussian model and resistor networks. 

The bulk free energy term $f_{\text{bulk}}$ is nonuniversal as are the surface free energies $f_{1s}$ and $f_{2s}$ and the subleading correction terms $f_p$. In contrast, $f_{\text{corn}}$ are believed to be universal \cite{Cardy1988}. In the $\lim_{\rho \to \infty}$ and $\lim_{\rho \to 0}$ the value of  $f_0(\rho)$ is related to the conformal anomaly  $c$ and conformal weights of the underlying conformal theory \cite{Blote1986,Affleck}. 
Moreover, in a rectangular geometry on a plane with free boundary conditions the leading finite-size correction term $f_0$ contains both universal $f_\mathrm{univ}$ and non-universal $f_\mathrm{nonuniv}$ contributions \cite{Kleban}. Thus, the term $f_0$ can be written as $f_0 = f_\mathrm{univ} + f_\mathrm{nonuniv}$. The universal part $f_\mathrm{univ}$ of $f_0$, which depends only upon the shape of the system and not upon the underlying lattice, can be calculated by conformal field theory methods, and for rectangular geometry with free boundary conditions is given by  \cite{Kleban}
\begin{equation}
f_\mathrm{univ}=-\frac{c}{4}\ln{\left[\eta(\rho)\eta{(1/\rho)}\right]}.
\label{funiv}
\end{equation}
Here $\rho=M/N$ is the aspect ratio and $\eta(\rho)$ is the Dedekind $\eta$ function. In this formula, $c$ represents the central charge defining the universality class of the system.  This formula also applies for example to percolation theory \cite{KlebanZiff98}.   However, the nonuniversal part $f_\mathrm{nonuniv}$ of $f_0$, which is lattice-dependent, is not calculable via conformal field theory methods. In this paper we will derive the leading finite-size correction term $f_0$ for the dimer model on rectangular $2M \times 2N$ lattice with free boundary conditions and explain why $f_0$  is different from Eq.\  (\ref{funiv}). We are also especially interested in the universal corner terms $f_{\text{corn}}$ because they are logarithmic. Cardy and Peschel studied the free energy within CFT \cite{Cardy1988} and  predicted that a corner with an angle $\pi/2$ and two edges under free boundary conditions has 
\begin{equation}
f_{\text{corn}}(0,0) = -\frac{c}{32}. \label{corner}
\end{equation} 
We confirmed this in Refs.\ \cite{Izmailian2012,Izmailian2013} for the Ising model on the square and triangular lattices with free boundary conditions. Later on, Imamura et al.\ \cite{Imamura} and Bondesan et al.\ \cite{Bondesan2012,Bondesan2013} also used CFT to study the corner terms with different conformally invariant boundary conditions and found that the contribution to the free energy from a corner with two edges under $a$ and $b$ boundary conditions is 
\begin{equation}
f_{\text{corn}}(a,b) = \Delta_{a,b} - \frac{c}{32}.
\label{fcornab} 
\end{equation}
This formula,  where $\Delta_{a,b}$ represents the conformal weight of the boundary operator inserted at the corner, was verified in our previous work on the Ising model on the square lattice  with different boundary conditions \cite{Izmailian2015c}. The contribution to the free energy from all corners $f_{\text{corn}}$ is given by
\begin{equation}
f_{\text{corn}}=\sum_if_{\text{corn}}(a_i, b_i) \label{fcorntotal}
\end{equation}
where summation is taken over all corners of the lattice and $a_i$ and $b_i$ denote boundary conditions on two edges of the corner $i$.

In this paper we will explain why the corner term $f_{\text{corn}}$ is absent in the asymptotic expansion of the free energy for the dimer model on a $2M \times 2N$ rectangular lattice with free boundary conditions.

The subleading correction terms $f_p$ in the asymptotic expansion of the free energy have been obtained for many models, including the Ising  \cite{Izmailian2002,Izmailian2007a,Izmailian2012a,Salas,Izmailian2002a} and  dimer models  \cite{Izmailian2002,Izmailian2003,Izmailian2014,Izmailian2006,Izmailian2016,Bleher2018} on different lattices under various boundary conditions, the Gaussian model \cite{Izmailian2002}, the spanning tree \cite{Izmailian2015a} and resistor networks \cite{Izmailian2014a,Izmailian2010}. In all these papers the subleading correction terms $f_p$ (for all p) are expressed in terms of Kronecker's double series, which in turn are directly related to elliptic theta functions. While the  Kronecker's double series can be expressed in terms of elliptic theta functions for arbitrary $p$, in fact in a series of papers \cite{Salas,Izmailian2002,Izmailian2003,Izmailian2006,Izmailian2007a,Izmailian2010,Izmailian2012a,Izmailian2014,Izmailian2014a,Izmailian2015a,Izmailian2016,Bleher2018} the exact expressions for the subleading correction terms $f_p$ for different models have been calculated up to second order (for $p=0,1,2$).  In  \cite{Izmailian2002a}, the subleading correction terms $f_p$ are calculated for the Ising model on a rectangular lattice under Brascamp-Kunz boundary conditions up to fourth order (for $p=0,1,2,3,4$).

In recent years, the dimer problem has continued to attract much interest, including the study of the asymptotic behavior.  Some references include 
\cite{Bleher2018,Pearce2017,Chhita2016,Morin2016,Kenyon2016,Allegra2015,Allegra2014,Nigro2018,Ghosh2007,Kenyon2006}.

In this paper we have calculated the exact expressions for the subleading correction terms $f_p$ in the asymptotic expansion of the free energy for the dimer model on a $2M \times 2N$  rectangular lattice with free boundary conditions explicitly up to 22nd order (for $p=0,1,2,...,22$), and also for the strip geometry.  Surprisingly we find that ratio of coefficients $f_p$ in the free energy expansion for the strip  ($f_p^\mathrm{strip}$)  and square ($f_p^\mathrm{sq}$)  geometry	$r_p=f_p^\mathrm{strip}/f_p^\mathrm{sq}$, in the limit of large $p$, tends to $1/2$ and predict that that ratio is exactly equal to 1/2 in the limit of $p \to \infty$.   Furthermore, we predict that the ratio of coefficients $f_p$ in the free energy expansion for rectangular ($f_p(\rho)$) for $\rho > 1$ and square ($f_p^\mathrm{sq}$)  geometry, multiplied by $\rho^{-p-1}$, that is $r_p={\rho^{-p-1} f_p(\rho)}/{f_p^\mathrm{sq}}$, is equal to $1/2$ in the limit of $p \to \infty$.
 

\section{Dimer model on rectangular lattice}
\label{introduction}
 
Let us consider the dimer model on the rectangular lattice $L$ of size $2M \times 2N$ of $4 M N$ sites with $2 M$ rows and $2 N$ columns, and with free boundary conditions. The partition function of the dimer model is given by
\begin{equation}
Z_{2M,2N}=\sum z_v^{n_v} z_h^{n_h}
\label{Zdimer}
\end{equation}
where the summation is taken over all possible dimer covering configurations, $z_h$ and $z_v$ are, respectively, the dimer weight in the horizontal and vertical directions, and $n_v$ and $n_h$ are, respectively, the number of vertical and horizontal dimers. 

In \cite{Izmailian2003} it was shown that the partition function of the dimer model on rectangular $2 M \times 2 N$ lattice under free boundary conditions, Eq.\ (\ref{Kasteleyn}), can be expressed in terms of $Z_{\frac{1}{2},\frac{1}{2}}(z,K,L)$, 
\begin{eqnarray}
Z_{2M,2N}^\mathrm{free}(z)&=&z_v^{2MN}\left[\frac{(1+z^2)^{\frac{1}{2}}\;
	Z_{\frac{1}{2},\frac{1}{2}}(z,2M+1,2N+1)}
{2z^{2N+1}\cosh{\left[(2M+1){\rm arcsinh}\,z\right]}
	\cosh{\left[(2N+1) {\rm arcsinh}\,\frac{1}{z}\right]}}\right]^{\frac{1}{2}},
\label{free}
\end{eqnarray}
where $z=z_h/z_v$ and $Z_{\frac{1}{2},\frac{1}{2}}(z,\cal M, \cal N)$ is given by
\begin{equation}
Z^2_{\frac{1}{2},\frac{1}{2}}(z,{\cal M},{\cal N})=
\prod_{m=0}^{{\cal M}-1}\prod_{n=0}^{{\cal N}-1}4\left(z^2 \sin^2{\frac{\pi\left(m+\frac{1}{2}\right)}{\cal M}}+\sin^2{\frac{\pi\left(n+\frac{1}{2}\right)}{\cal N}}\right). \label{Zab}
\end{equation}
For isotropic case ($z_v=z_h=z=1$), Eq.\  (\ref{free}) reduces to
\begin{eqnarray}
Z_{2M,2N}^\mathrm{free}&=&\left[\frac{
	Z_{\frac{1}{2},\frac{1}{2}}(2M+1,2N+1)}
{\sqrt{2}\,\cosh{\left[(2M+1){\rm arcsinh}\,1\right]}
	\cosh{\left[(2N+1) {\rm arcsinh}\,1\right]}}\right]^{\frac{1}{2}}
\label{free1}
\end{eqnarray}
where arcsinh $1 = \ln(1+\sqrt{2})$.  This is an alternate expression for $Z$ of Eq.\ (\ref{Kasteleyn}).  Based on such results, one can easily write down all the terms of the exact asymptotic expansion of the logarithm of the partition functions for the dimer model using an asymptotic expansion of $\ln Z_{\frac{1}{2},\frac{1}{2}}(z,2M+1,2N+1)$ (see \cite{Izmailian2002,Izmailian2003})
\begin{equation}
\ln Z_{\frac{1}{2},\frac{1}{2}}(z,2M+1,2N+1)=\frac{S}{\pi}\int_0^{\pi}
\omega_{z}(x)dx+
\ln\frac{\theta_3(z\rho)}
{\eta(z\rho)}-2\pi\rho\sum_{p=1}^\infty
\left(\frac{\pi^2 \rho}{S}\right)^p \frac{\Lambda_{2p}}{(2p)!}
\frac{K_{2p+2}^{\frac{1}{2},\frac{1}{2}}(z \rho)}{2p+2} 
\label{ExpansZab}
\end{equation}
where $S =(2M+1)(2N+1)$ is the area of the lattice with an additional row and column,  $\rho=(2M+1)/(2N+1)$ is the aspect ratio,  
$\eta(\tau)=e^{-\pi \tau /12}\prod_{n=1}^\infty\left[
1-e^{- 2\pi \tau n}\right]$
is the Dedekind $\eta$ function,
and $K_{2p+2}^{\frac{1}{2},\frac{1}{2}}(\tau)$ is a special case of Kronecker's double series \cite{Izmailian2002}, defined by
\begin{equation}
K_{p}^{\alpha,\beta}(\tau)= - \frac{p!}{(-2 \pi i)^p} \sum_{(m,n) \ne (0,0)} \frac{e^{-2 \pi i (n \alpha + m \beta)}}{(n +\tau m)^p}
\end{equation}
The functions $\theta_2(\tau), \theta_3(\tau), \theta_4(\tau)$ are elliptic theta functions with nome $q=e^{-\pi \tau}$, $\tau$ being real here. The Dedekind $\eta$ function can be expressed as
\begin{equation}
\eta(\tau)=\left[\theta_{2}( \tau)\theta_{3}( \tau)\theta_{4}( \tau)/2\right]^{1/3}. \label{eta}
\end{equation}
Also in Eq.\ (\ref{ExpansZab}), we have
\begin{equation}
\omega_{z}(k)={\rm arcsinh}\left(z \sin{k}\right).
\label{omega}
\end{equation}
The differential operators $\Lambda_{2p}$ that appear in Eq.\  (\ref{ExpansZab}) can be expressed via coefficients $z_{2p}$ of the Taylor expansion of the lattice dispersion relation $\omega_{z}(k)$ and derivatives over $z$ (see Appendix \ref{Lambda}).
To get result for the dimer model in the isotropic case we should evaluate Eqs.\ (\ref{ExpansZab}) and (\ref{omega}) at $z=1$.
From Eqs.\ (\ref{free1}) and (\ref{ExpansZab}) one can easily find  the exact asymptotic expansion of the free energy $f=\frac{1}{S}\ln Z_{2M,2N}^\mathrm{free}$ for the dimer model in the isotropic case ($z=1$), which can be written as
\begin{equation}
f=\frac{1}{S}\ln Z_{2M,2N}^\mathrm{free}=f_\mathrm{bulk}+\frac{2 f_{1s}}{2N+1}+\frac{2f_{2s}}{2M+1}
+\frac{f_0(\rho)}{S}+
\sum_{p=1}^\infty f_{p}(\rho)S^{-p-1},
\label{expansion}
\end{equation}
The bulk free energy $f_\mathrm{bulk}$ is  given by
\begin{equation}
f_\mathrm{bulk}=\frac{1}{2\pi}\int_0^\pi\omega_1(x)dx=\frac{1}{2\pi}\int_0^\pi{\rm arcsinh}{\left( \sin{x}\right)}dx=\frac{G}{\pi},
\label{fbulk}
\end{equation}
where $G$ is the Catalan constant given by
$G=\sum_{n=0}^{\infty}{(-1)^n}/{(2n+1)^2}=0.915 965 594 \dots$.
The surface free energies $f_{1s}$ and $f_{2s}$ are given by
\begin{equation}
f_{1s}=f_{2s}=-\frac{1}{4}\ln(1+\sqrt{2})\label{fs}
\end{equation}
We will show that all finite-size correction terms $f_p(\rho)$ for $p=0,1,2,...$ are invariant under the transformation
\begin{equation}
\rho \to 1/\rho \label{r1r}
\end{equation}
as one would expect by the symmetry of the rectangular system.
The leading coefficient $f_0$ is given by
\begin{equation}
f_0=\frac{3}{4}\ln 2+\frac{1}{2}\ln\frac{\theta_3(\rho)}
{\eta(\rho)}.\label{f0dimer}
\end{equation}
It is easy to see from Eqs.\ (\ref{theta3}) and (\ref{eta1})  (see Appendix \ref{Jacobi}) that $f_0(\rho)$ is invariant under transformation  given by Eq.\  (\ref{r1r})
\begin{equation}
f_0\left(\frac{1}{\rho}\right)=f_0(\rho)\label{Jacobif0}.
\end{equation}
The coefficients $f_p(\rho)$ are given by
\begin{equation}
f_p(\rho)=-\frac{\pi^{2p+1}\rho^{p+1}}{(2p)!(2p+2)}
\left[\Lambda_{2p}\,K_{2p+2}^{\frac{1}{2},\frac{1}{2}}(z \rho)\right]_{z=1}
\label{fpoocc}
\end{equation}
Again, it is easy to see from Eq.\  (\ref{Kronecker})  that $f_p(\rho)$ is invariant under transformation given by Eq.\  (\ref{r1r}):
\begin{equation}
f_p\left(\frac{1}{\rho}\right)=f_p(\rho)\label{Jacobifp}
\end{equation}
Comparing Eqs.\ (\ref{freeenergy}) and (\ref{expansion}) one can see that there are no corner terms in the asymptotic expansion of the free energy (Eq.\  (\ref{expansion})) for the dimer model $2M \times 2N$ lattice with free boundary conditions, while the rectangular lattice with free boundary conditions has four corners. One can explain that paradox if we consider the bijection between dimer coverings, spanning trees and sandpile models. 
\begin{figure}[!ht]
  	\includegraphics[width=150mm]{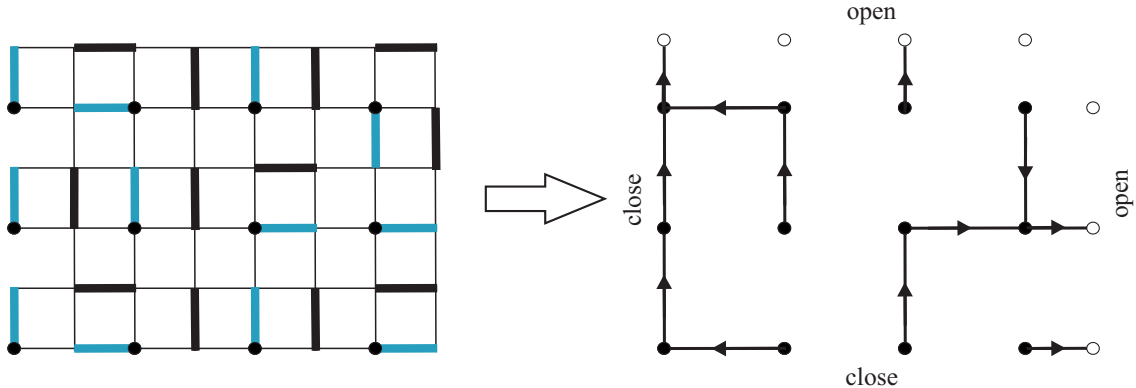}
	\caption{Mapping of a dimer covering on lattice $L$ to a spanning tree on the odd-odd sublattice $G$. The dimer covering on a $6 \times 8$ lattice $L$ is shown on the left, where the dimers which touch the odd-odd sublattice G are coloured in blue, and the lower left-hand corner has coordinates (1,1). The right figure shows the corresponding spanning tree on the $3 \times 4$ sublattice $G$. The solid dots represent the sites of the odd-odd sublattice $G$, and the open dots are the roots of the trees.}
	\label{steps1}
\end{figure}

The bijection between  close-packed dimer coverings of lattice $L$ and spanning trees on the odd-odd sublattice $G \subset L$ is well known \cite{Temperley,Izmailian2005,Izmailian2007}. The odd-odd sublattice $G \subset L$ contains the sites whose coordinates are both odd. A dimer containing a site of $G$, in blue in Fig.\ \ref{steps1}, can be represented as an arrow directed along the dimer from the odd-odd site to the nearest neighbor site of $G$. It is easy to prove that the resulting set of arrows generates a uniquely defined spanning tree (see Fig.\ \ref{steps1}). Since the dimers which do not contain a site of $G$ are completely fixed by the others, one has a one-to-one correspondence between dimer coverings on $L$ and spanning trees on $G$.  
The above construction leads to a set of spanning trees on the odd-odd sublattice $G$, where certain arrows may point out of the lattice from the right vertical side and upper horizontal side as in Fig.\ \ref{steps1}. Viewing these vertical and horizontal boundaries of $G$ as roots for the spanning trees, we see that dimer coverings on $L$ map onto spanning trees on $G$ which can grow from any site of the right and upper sides of the lattice $G$. The sites on these
boundaries, being connected to roots, are dissipative in the Abelian Sandpile Model (ASM) language, and form open boundaries. Thus the spanning trees map onto the ASM  configurations with one vertical open and one horizontal open dissipative boundary, and the two other closed boundaries.

Thus also the dimer model on rectangular $2M \times 2N$ lattices with free boundary conditions, which maps to spanning trees, also maps onto the ASM configurations with the one vertical and one horizontal open dissipative boundary, and the two closed boundaries. Now the contribution to the free energy from all four corners $f_{\text{corn}}$ of the lattice is given by Eq.\  (\ref{fcorntotal}), where the summation is taken over four corners of the lattice. As one can see from Fig.\ (\ref{steps1}), we have two corners with boundary conditions on the two sides of the corner that are the same (open-open and closed-closed), and two corners with boundary conditions on the two sides of the corner that are different (open-closed and closed-open).

When boundary conditions on both sides of the corner are the same, the boundary operator inserted at the corner is just the identity operator with $\Delta_{aa} = 0$. In the case when the boundary condition on one side of the corner is $a_i$ = closed and on the other side it is $b_i$ = open, the boundary operator at the corner,  which changes the boundary conditions from open to closed or closed to open, has conformal weight $\Delta_\mathrm{open,closed}=\Delta_\mathrm{closed,open} = -\frac{1}{8}$ \cite{Ruelle}.  Thus for $f_{\text{corn}}$ we obtain from Eq.\  (\ref{fcorntotal}) that
\begin{equation}
f_{\text{corn}}=\sum_{i=1}^4\left(\Delta_{a_i,b_i} - \frac{c}{32}\right)=-2\frac{c}{32}+2\left(-\frac{1}{8} - \frac{c}{32}\right)=-\frac{1}{4}-\frac{c}{8} \label{fcornc2}
\end{equation}
The central charge for the dimer model on rectangular lattice is known to be $c=-2$ \cite{Izmailian2005,Izmailian2007}. Now plugging $c=-2$ to Eq.\  (\ref{fcornc2}) we obtain that the contribution to the free energy from all four corners $f_{\text{corn}}$ is equal to zero
\begin{equation}
f_{\text{corn}}=0 \label{fcornc2a}
\end{equation}
which is in full agreement with our results for the exact asymptotic expansion of the free energy for the dimer model on a plane, Eq.\  (\ref{expansion}). 

The leading coefficient $f_0$ is given by Eq.\  (\ref{f0dimer}). Comparing Eqs.\ (\ref{funiv}) and  (\ref{f0dimer}) we can see clear a difference between our results (see Eq.\  (\ref{f0dimer})) and result for the universal part $f_\mathrm{univ}$ of $f_0$ in the rectangular geometry with free boundary conditions \cite{Kleban} (see Eq.\  (\ref{funiv})). Again we can explain that paradox by considering the bijection between dimer coverings, spanning trees and the sandpile model. As we show above the dimer model on the rectangular $2M \times 2N$ lattice with free boundary conditions can be mapped onto the ASM configurations with two open and two closed boundaries as shown in Fig.\ \ref{steps1}. Thus our result for $f_0$ is related to the rectangular $2M \times 2N$ lattice  with two open and two closed boundaries and is different from the universal part $f_\mathrm{univ}$, which is calculated by conformal field theory methods for rectangular geometry with free boundary conditions on all sides (see Eq.\  (\ref{funiv})).

The subleading coefficients $f_p(\rho)$ for $p=1,2,3,...$ are given by Eq.\  (\ref{fpoocc}). Now using expressions for $K_{2p+2}^{\frac{1}{2},\frac{1}{2}}(\rho)$ and $\Lambda_{2p}$, which are given in Appendices \ref{Lambda} and \ref{KroneckerToTheta}, and the following relations between the elliptic functions and  derivatives of the elliptic functions
\begin{eqnarray}
\frac{\partial}{\partial z}\ln{{\theta}_3} &=&
\frac{\pi}{4}{\theta}_4^4+\frac{\partial}{\partial z}\ln{{\theta}_2} \nonumber\\
\frac{\partial}{\partial z}\ln{{\theta}_4} &=&
\frac{\pi}{4}{\theta}_3^4+\frac{\partial}{\partial z}\ln{{\theta}_2}\nonumber\\
\frac{\partial}{\partial z}\ln{\theta_2} &=& -\frac{1}{2}{\theta}_3^2 E,\nonumber\\
\frac{\partial E}{\partial z}
&=&\frac{\pi^2}{4}{\theta}_3^2{\theta}_4^4-\frac{\pi}{2}{\theta}_4^4 E\nonumber
\end{eqnarray}
we can express the subleading correction terms $f_p(\rho)$ in the asymptotic expansion of the free energy for the dimer model on a $2M \times 2N$ rectangular lattice with free boundary conditions for any value of $p$ in terms of the elliptic theta functions ${\theta}_2, {\theta}_3, {\theta}_4$ and the elliptic integral of the second kind $E$. In particular in this paper we have calculated the subleading correction terms $f_p$ in terms of the elliptic functions the elliptic integral of the second kind up to $p=22$. Due to very large expressions for $f_p(\rho)$ for $p > 5$ we have not listed the expressions of $f_p(\rho)$ for $p>4$ here, instead we have listed the exact values for $f_p(\rho)$ for particular value of $\rho$, namely $\rho =1$ and $\rho =2$, up to $p=22$, in Table \ref{tabular}.

The elliptic theta functions and the elliptic integral of the second kind $E$ at particular values of the aspect ratio $\rho =1$ and $\rho =2$ are given by
\begin{eqnarray}
\theta_2&=&\theta_4=\frac{(\pi/2)^{1/4}}{\Gamma(3/4)}, \quad \theta_3=\frac{\pi^{1/4}}{\Gamma(3/4)}, \quad E=\frac{\pi^{3/2}}{4\, \Gamma(3/4)^2} + \frac{\Gamma(3/4)^2}{2\sqrt{\pi}}   \label{rho1}
 \end{eqnarray}   
 for $\rho = 1$ and 
 \begin{eqnarray}
\theta_2&=&\frac{\pi^{1/4}}{\Gamma(3/4)\sqrt{2(2+\sqrt{2})}}, \quad \theta_3=\frac{\pi^{1/4}\sqrt{2+\sqrt{2}}}{2\,\Gamma(3/4)}, \quad \theta_4=\frac{\pi^{1/4}}{2^{1/8}\Gamma(3/4)}, \nonumber\\
E&=&\frac{\pi^{3/2}}{2\sqrt{2}\, \Gamma(3/4)^2} + \frac{\Gamma(3/4)^2}{(2+\sqrt{2})\sqrt{\pi}}  \label{rho2}
\end{eqnarray}
for $\rho = 2$, where $\Gamma(z)$ is the gamma function.
Now, using the above expressions and the expressions for subleading correction terms $f_p(\rho)$ in terms of the elliptic theta functions and the elliptic integral, we obtain the exact values for $f_p(\rho)$  for $\rho =1$ and $\rho =2$ up to $p=22$ (see Supplementary Materials). Numerical values of $f_p(\rho)$ at particular values of the aspect ratio $\rho$, namely  $\rho =1$ and $\rho =2$ are given in Table \ref{tabular}. 
The difference between exact  $f^\mathrm{exact}(N)$ and asymptotic value $f_{p_{\mathrm{max}}}^\mathrm{asympt}(N)$ (for $p_{\mathrm{max}}=2$ and $10$) of the free energy for dimer model on the square lattice with side $2N$  are given in Table \ref{tab}. Here $p_{max}$ is a cut off of the asymptotic series, Eq.\ (\ref{expansion}).

\begin{figure}[!ht]
	\includegraphics[width=150mm]{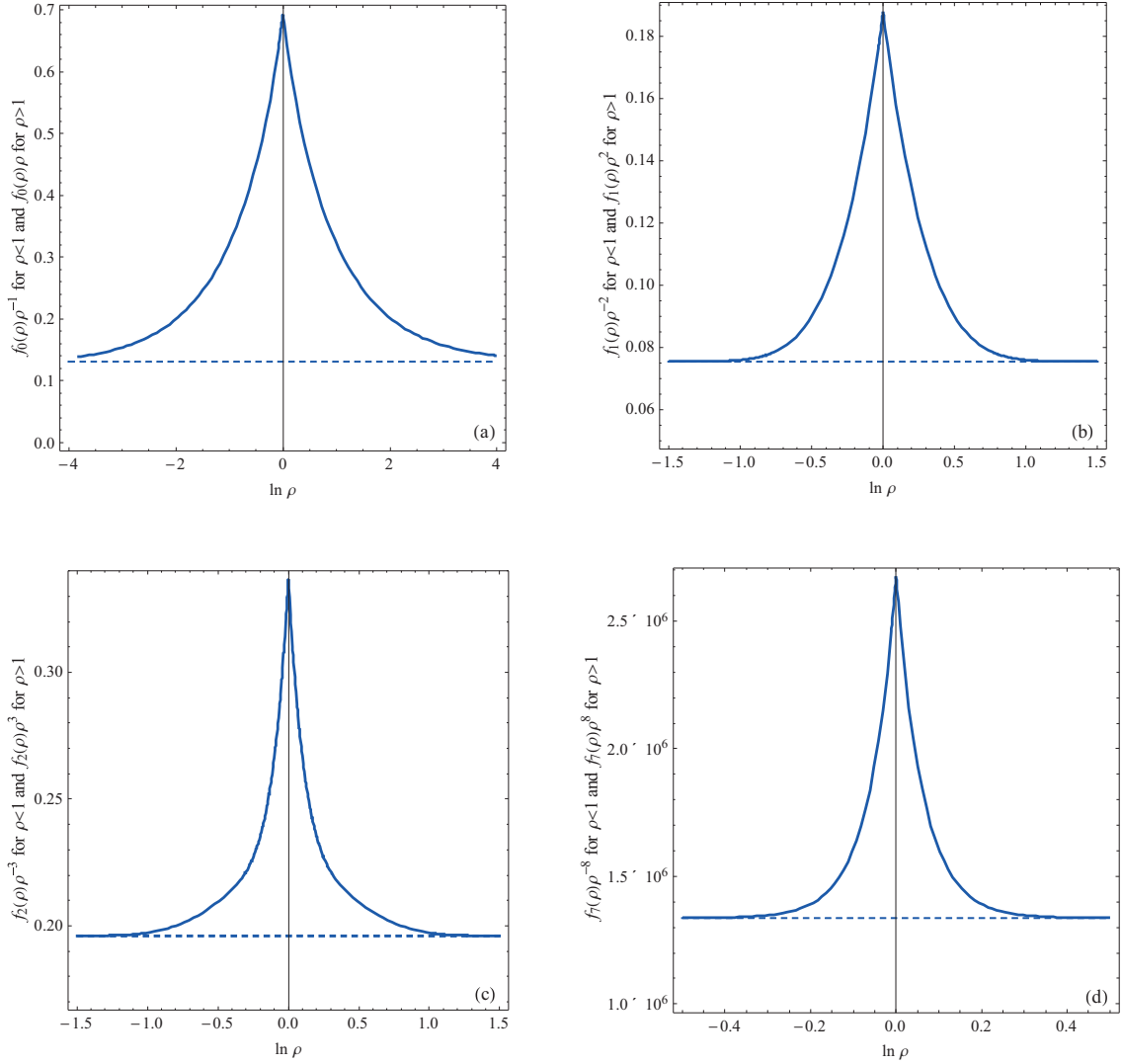}
	\caption{The behavior of the correction
		terms $f_p(\rho)$ for (a) $p=0$, (b) $p=1$, (c) $p=2$, (d) $p=7$. The dotted
		lines represent the corresponding value of $f_p$ in strip geometry.}
	\label{fig2}
\end{figure}

In Fig.\ (\ref{fig2}) we plot the behavior of the subleading correction terms $\rho^{\pm (p + 1)}f_p(\rho)$  ($+$ for $\rho > 1$ and $-$ for $\rho < 1$) for (a) $p=0$, (b) $p=1$, (c) $p=2$, (d) $p=7$ as a function of the aspect ratio $\rho$.

\begin{table}[ht]
	\caption{Coefficients $f_k$ in the asymptotic expansion of the free energy, for the square, rectangle of aspect ratio 2, and the infinite strip.}
	\label{tabular}
	\begin{center}
		\begin{tabular}{| c | c | c | c |}
			\hline
			$p$ & $f_{p}^\mathrm{sq}(1)$ & $f_{p}(2)$ & $f_{p}^\mathrm{strip}$ \\\hline
			0 & 0.693147180559... & 0.783525480959... & 0.130899693899... \\\hline
			1 & 0.188074112577... & 0.320716568887... & 0.075362478042... \\\hline
			2 & 0.336874969732... & 1.620602011573... & 0.196069159812... \\\hline
			3 & 3.290053550462... & $2.50926174993...\times 10^1$ & 1.565734641321... \\\hline
			4 & $4.81457307505... \times10^1$ & $7.84426067442...\times 10^2$ & $2.44962573304...\times 10^1$ \\\hline
			5 & $1.27811955024...\times 10^3$ & $4.06639855884...\times 10^4$ & $6.35177661759...\times 10^2$ \\\hline
			6 & $4.91797597262...\times 10^4$ & $3.15446270980...\times 10^6$ & $2.46425620918...\times 10^4$ \\\hline
			7 & $2.67530735325...\times 10^6$ & $3.42195018099...\times 10^8$ & $1.33668213673...\times 10^6$ \\\hline
			8 & $1.93146021814...\times 10^8$ & $4.94576151005...\times 10^{10}$ & $9.65965464746...\times 10^7$ \\\hline
			9 & $1.79421498557...\times 10^{10}$ & $9.18564336803...\times 10^{12}$ & $8.97034559509...\times 10^9$ \\\hline
			10 &$2.08172298698...\times 10^{12}$ & $2.13174241269...\times 10^{15}$ & $1.04088960155...\times 10^{12}$ \\\hline
			11 & $2.95165319167...\times 10^{14}$ & $6.04493217496...\times 10^{17}$ & $1.47581343488...\times 10^{14}$ \\\hline
			12 & $5.02086224796...\times 10^{16}$ & $2.05655123614...\times 10^{20}$ & $2.51043848305...\times 10^{16}$ \\\hline
			13 & $1.00920433549...\times 10^{19}$ & $8.26739400791...\times 10^{22}$ & $5.04601683138...\times 10^{18}$ \\\hline
			14 & $2.36629755293...\times 10^{21}$ & $3.87694313074...\times 10^{25}$ & $1.18314914769...\times 10^{21}$ \\\hline
			15 & $6.40123719928...\times 10^{23}$ & $2.09755719148...\times 10^{28}$ & $3.20061827240...\times 10^{23}$ \\\hline
			16 & $1.97885518164...\times 10^{26}$ & $1.29686257503...\times 10^{31}$ & $9.89427623712...\times 10^{25}$ \\\hline
			17 & $6.93253270671...\times 10^{28}$ & $9.08660917146...\times 10^{33}$ & $3.46626631597...\times 10^{28}$ \\\hline
			18 & $2.73210457870...\times 10^{31}$ & $7.16204825184...\times 10^{36}$ & $1.36605229412...\times 10^{31}$ \\\hline
			19 & $1.20333420978...\times 10^{34}$ & $6.30893685465...\times 10^{39}$ & $6.01667104210...\times 10^{33}$ \\\hline
			20 & $5.88863561006...\times 10^{36}$ & $6.17468197571...\times 10^{42}$ & $2.94431780611...\times 10^{36}$ \\\hline
			21 & $3.18488175869...\times 10^{39}$ & $6.67918114922...\times 10^{45}$ & $1.59244087916...\times 10^{39}$ \\\hline
			22 & $1.89474701875\dots\times 10^{42}$ & $7.94714500005...\times 10^{48}$ & $9.47373509413...\times 10^{41}$ \\\hline
		\end{tabular}
	\end{center}
\end{table}

\begin{table}[ht]
	\caption{The exact free energy $f^\mathrm{exact}(N)$ for a square with side $2N$ calculated from Eq.\ (\ref{Kasteleyn}), and the difference between exact and asymptotic values of the free energy $f_{p_\mathrm{max}}^\mathrm{asympt}(N)$ (for $p_\mathrm{max}=2$ and $10$), where $p_\mathrm{max}$ is a cut-off of the asymptotic series, Eq.\ (\ref{expansion}).}
	\label{tab}
	\begin{center}
		\begin{tabular}{| c | c | c | c |}
			\hline
			$N$ & $f^\mathrm{exact}(N)$ & $f^\mathrm{exact}(N)-f_{2}^\mathrm{asympt}(N)$ & $f^\mathrm{exact}(N)-f_{10}^\mathrm{asympt}(N)$  \\\hline
			2 & 0.14334075753824440006 & $6.205110750\times10^{-6}$ & -0.00114988029 \\\hline
			4 & 0.20221727453649946996 & $9.710971517\times10^{-8}$ & $ -6.123675796 \times10^{-9}$ \\\hline
			8 & 0.24211609833554829981 & $ 4.980854240\times10^{-10}$ & $ 1.777260867\times10^{-15}$ \\\hline
			16 & 0.26548927115575958947 & $ 2.371559964\times10^{-12}$ & $ 1.266298133\times10^{-22}$ \\\hline
			32 & 0.27816537942911183803 & $ 1.036113289\times10^{-14}$ & $ 9.508968901\times10^{-30}$ \\\hline
			64 & 0.28477020429871780721 &  $ 4.294055324\times10^{-17}$& $ 6.612762353\times10^{-37}$ \\\hline
			128 & 0.28814192930242180726 & $ 1.729156214\times10^{-19}$ & $ 4.293295929\times10^{-44}$ \\\hline
			256 & 0.28984546071399267489 & $ 6.859432660\times10^{-22}$ & $ 2.676266774\times10^{-51}$ \\\hline
		\end{tabular}
	\end{center}
\end{table}
\section{Dimer model on an infinitely long strip}
\label{introduction1}
Let us consider the case of an infinitely long strip ($M \to \infty$), with $N$ fixed. The free energy per site ($f^\mathrm{strip}(N)$) for that case can be written as
\begin{eqnarray}
f^\mathrm{strip}(N)=\lim_{M \to \infty}\frac{1}{S}\ln Z_{2M,2N}^\mathrm{free}=-\frac{{\rm arcsinh}\,1}{2(2N+1)}+\lim_{M \to \infty}\frac{1}{2S} \ln Z_{
	\frac{1}{2},\frac{1}{2}
}(2M+1,2N+1)
\label{free2}
\end{eqnarray}
where $S=(2M+1)(2N+1)$ is the area.
In the case of an infinitely long strip ($M, \rho \to \infty$), where $\rho=(2M+1)/(2N+1)$, we have
\begin{eqnarray}
\lim_{M \to \infty}\theta_{3}(\rho)&=&1\\
\lim_{M \to \infty}\eta(\rho)&=&\lim_{\rho \to \infty}e^{-\pi \rho /12}=0\\
\lim_{M \to \infty}\Lambda_{2p}K_{2p+2}^{\frac{1}{2},\frac{1}{2}}(\rho )&=&z_{2p}B_{2p+2}^{1/2}
\end{eqnarray}
where $B_{2p+2}^{1/2}$ are Bernoulli  polynomials $B_n(z)$ evaluated at $z = 1/2$: ($B_{2}^{1/2}=-{1}/{12}, B_{4}^{1/2}={7}/{240}, B_{6}^{1/2}=-{31}/{1344}, B_{8}^{1/2}={127}/{3840}, B_{10}^{1/2}=-{2555}/{33792},...$) and Eq.\  (\ref{ExpansZab}) can be rewritten as
\begin{eqnarray}
f^\mathrm{strip}(N)&=&-\frac{{\rm arcsinh}\,1}{2(2N+1)}+\lim_{M \to \infty}\frac{1}{2S}\ln Z_{\frac{1}{2},\frac{1}{2}}(2M+1,2N+1)\nonumber\\
&=&\frac{2G}{\pi}-\frac{{\rm arcsinh}\,1}{2(2N+1)}+
\frac{\pi}
{12\,(2N+1)^{2}}-\sum_{p=1}^\infty
\frac{2\pi^{2p+1}}{(2N+1)^{2p+2}}\frac{z_{2p}}{(2p)!}
\frac{B_{2p+2}^{1/2}}{2p+2},
\label{ExpansZab1}
\end{eqnarray}
Thus the exact asymptotic
expansion of the free energy per site for the dimer model on an infinitely long strip can be written as
\begin{equation}
f^\mathrm{strip}(N)=f_\mathrm{bulk} +\frac{2 f_s}{2N+1} + \sum_{p=0}^\infty \frac{f_{p}^\mathrm{strip}}{(2N+1)^{2p+2}},
\label{expansion1}
\end{equation}

The bulk free energy $f_\mathrm{bulk}$ is  given by
\begin{eqnarray}
f_\mathrm{bulk}&=&\frac{G}{\pi},
\nonumber
\end{eqnarray}
where $G$ is the Catalan constant. The surface free energy $f_{s}$ is given by
\begin{equation}
f_s=f_{1s}=-\frac{1}{4}\ln(1+\sqrt{2})\label{fs1}
\end{equation}
and the coefficients $f_p^\mathrm{strip}$ are given by
$$
f_p^\mathrm{strip}=-\frac{\pi^{2p+1}}{(2p)!(2p+2)}
z_{2p}\,B_{2p+2}^{\frac{1}{2}}.
$$
The leading $f_0^\mathrm{strip}$ is given by
\begin{equation}
f_0^\mathrm{strip}=\frac{\pi}{24}=0.1308996938995747...
\label{f0strip}
\end{equation}
For reader convenience we list subleading coefficients $f_p^\mathrm{strip}$ up to 22nd order in Supplementary Materials.

On the basis of conformal invariance, the asymptotic finite-size scaling behavior of the free energy per site of an infinitely long strip of finite width ${\cal N}$ at criticality is expected to have  the form \cite{Blote1986,Affleck}
\begin{equation}
f=f_\mathrm{bulk}+\frac{2f_s}{{\cal N}}+\frac{f_0}{{\cal N}^2}+...
\label{stripconf}
\end{equation}
where ${\cal N}$ equal $2 N + 1$ in our problem.
The value of $f_0$ is
related to the conformal charge $c$ of the underlying conformal theory and depends on the boundary conditions in the transversal direction; in the strip geometry it is given by
\begin{equation}
f_0=\pi\left(\frac{c}{24}-\Delta\right)\label{f0conf}
\end{equation}
where the number $\Delta$ is the conformal weight of the
operator with the smallest scaling dimension present in the
spectrum of the Hamiltonian with the given boundary
conditions. 
For the dimer model on a rectangular lattice the central charge is $c=-2$ \cite{Izmailian2005,Izmailian2007} and the conformal weight of the operator with open-closed boundary conditions is equal to $\Delta=-1/8$. Plugging these values of $c$ and $\Delta$ to Eq.\  (\ref{f0conf}) we obtain
$$
f_0=\frac{\pi}{24}
$$
in full agreement with our result (see Eq.\  (\ref{f0strip})). Once again we confirm that the dimer model on $2M \times 2N$ rectangular lattice with free boundary conditions should be considered as model with two open and two closed boundary conditions as shown in Fig.\ \ref{steps1}.
   
It is easy to show that 
\begin{equation}
f_p^\mathrm{strip}=\lim_{\rho \to \infty}f_p(\rho)\rho^{-p-1}\label{fpstr}
\end{equation}
For let us consider the exact asymptotic expansion  of the free energy per site for the dimer model on $2M \times 2N$ rectangular lattice on plane (see Eq.\  (\ref{expansion})). Taking the limit $M \to \infty$ one can obtain the exact asymptotic expansion of the free energy per site for the dimer model on an infinitely long strip
\begin{equation}
f^\mathrm{strip}(N)=\lim_{M \to \infty}\frac{1}{S}\ln Z_{2M,2N}^\mathrm{free}=f_\mathrm{bulk} +\frac{f_{1s}}{2N+1} + \lim_{M \to \infty}\sum_{p=0}^\infty f_{p}(\rho)S^{-p-1},
\label{expansion2}
\end{equation}
Here $S=(2M+1)(2N+1)$ can be rewritten as $S=\rho (2N+1)^2$, where $\rho=(2M+1)/(2N+1)$ is the aspect ratio. Plugging $S=\rho (2N+1)^2$ back to Eq.\  (\ref{expansion2}) we obtain
\begin{equation}
f^\mathrm{strip}(N)=\lim_{M \to \infty}\frac{1}{S}\ln Z_{2M,2N}^\mathrm{free}=f_\mathrm{bulk} +\frac{2f_{1s}}{2N+1} + 
\sum_{p=0}^\infty\frac{\lim_{\rho \to \infty}f_{p}(\rho)\rho^{-p-1}}{(2N+1)^{2p+2}},\label{expansion3}
\end{equation}
Now comparing Eqs.\ (\ref{expansion1}) and (\ref{expansion3}) we obtain Eq.\  (\ref{fpstr}).

\section{Exact results on the strip}

For the strip of width 2N, we can also derive a simple exact expression for $f^\mathrm{strip}(N)$ with a single sum. With the help of the identity 
\begin{equation}
\prod_{m=0}^{M-1}4\textstyle{\left[~\!{\rm sinh}^2\omega + \sin^2\left(\frac{\pi (m+1/2)}{M}\right)\right]}= 4\,{\rm cosh}^2\left(M\,\omega\right)
\end{equation} the partition function
the  $Z_{\frac{1}{2},\frac{1}{2}}(z,{\cal M}, {\cal N})$ given by Eq.\ (\ref{Zab}) can be transformed into simpler form
\begin{equation}
Z_{\frac{1}{2},\frac{1}{2}}(z,{\cal M}, {\cal N})=\prod_{n=0}^{{\cal N} -1} 2 {\rm cosh}\left[{\cal M}\omega_z \left(\frac{\pi(n+1/2)}{N}\right)\right] \label{Zpartab}
\end{equation}
where $\omega_z(k)$ is given by Eq.\ (\ref{omega}). In the limit ${\cal M} \to \infty$ from Eq.\ (\ref{Zpartab}) we can obtain  
\begin{eqnarray}
\lim_{{\cal M} \to \infty}\left[Z_{\frac{1}{2},\frac{1}{2}}(z,{\cal M}, {\cal N})\right]^{\frac{1}{{\cal M}}}&=&\prod_{n=0}^{{\cal N} -1} e^{\omega_z \left(\frac{\pi(n+1/2)}{N}\right)}=\prod_{n=0}^{{\cal N} -1} e^{{\rm arcsinh}\sin\left(\frac{\pi(n+1/2)}{N}\right)}\nonumber\\
&=&\prod_{n=0}^{{\cal N}-1}  \left( \sin \frac{\pi(n+1/2) }{{\cal N}}+ \sqrt{1+\sin^2\frac{\pi(n+1/2) }{{\cal N}}} \right) \label{Zpartab1}
\end{eqnarray}
Thus the free energy per site ($f^\mathrm{strip}(N)$) for infinitely long strip given by Eq.\ (\ref{free2}) can be further simplified as
\begin{eqnarray}
f^\mathrm{strip}(N)&=&\lim_{M \to \infty}\frac{1}{S}\ln Z_{2M,2N}^\mathrm{free}=-\frac{{\rm arcsinh}\,1}{2(2N+1)}+\lim_{M \to \infty}\frac{1}{2S} \ln Z_{\frac{1}{2},\frac{1}{2}}(2M+1,2N+1)\nonumber\\
&=&-\frac{{\rm arcsinh}\,1}{2(2N+1)}+\frac{1}{2(2 N + 1)} \sum_{n=0}^{2N} {\rm arcsinh}\left(\sin\frac{\pi(n+1/2)}{N}\right)\label{free21}\\
&=&\frac{1}{2 N + 1} \sum_{n=0}^{N-1} {\rm arcsinh}\left(\sin\frac{\pi(n+1/2)}{N}\right)\label{exactstrip}\\
&=&\frac{1}{2 N + 1} \sum_{n=0}^{N-1}\ln\left(\sin \frac{\pi(n+1/2) }{2N+1}+\sqrt{1+\sin^2 \frac{\pi(n+1/2) }{2N+1}} \right)\label{exactstrip1}
\end{eqnarray}
or equivalently
\begin{eqnarray}
Z^\mathrm{strip}(N) &=& e^{(2N+1)f^\mathrm{strip}(N)} = \prod_{n=0}^{N-1}  \left( \sin \frac{\pi(n+1/2) }{2 N+1}+ \sqrt{1+\sin^2\frac{\pi(n+1/2) }{2 N+1}} \right) 
\label{exactstripproduct}
\end{eqnarray}
In transition from Eq.\ (\ref{free21}) to Eq.\ (\ref{exactstrip}) we first split summation in Eq.\ (\ref{free21}) to two parts
$$
\sum_{n=0}^{2N} \to \sum_{n=0}^{N-1}+\sum_{n=N}^{2N}  
$$
and then in second sum we have changed variable $n \to 2N-n$. 

Note, that with the help of the Euler-Maclaurin summation formula (see, for example, Appendix A \cite{Izmailian2002}) one can easily find from Eq.\ (\ref{free21}) the asymptotic expansion of the free energy per site for the dimer model on an infinitely long strip given by Eq.\ (\ref{ExpansZab1}).

With Eq.\ (\ref{exactstrip}) one can easily calculate $f^\mathrm{strip}(N)$ to high precision for $N$ up to $10^6$.
In comparison, Eq.\ (\ref{expansion1}) represents an asymptotic series, which is non-convergent for a fixed $N$.  Typical of asymptotic series, the difference between successive terms decreases as $p$ 
increases, up to a point, and then starts increasing again, eventually exponentially.  The best approximation typically occurs when the contribution of the last term in the sum is at a minimum.
In this case, the coefficients of the asymptotic series are all positive, so the partial sums in Eq.\ (\ref{expansion}) are a monotonically increasing function of $p$.   In Table \ref{table:strip} we show exact values of $f$ calculated from Eq.\ (\ref{exactstrip}), and compare them with the results of the asymptotic series cut off when the last term is a minimum, to the precision of the last term.  As one can see, the exact value and asymptotic value truncated to the precision of the last term are in excellent agreement.   Similar behavior will occur for the asymptotic series for the rectangle, Eq.\ (\ref{expansion}).

The approach from a rectangle to a strip is also interesting to study.  From Eq.\ (\ref{expansion}) one can show that for $N$ fixed, and large but finite $M$, one has
\begin{equation}
f(N,M) \sim f^\mathrm{strip}(N) + \frac{a(N)}{2 M + 1}  
\label{largeM}
\end{equation}
where $a(N) = -\frac12 \ln (1 + \sqrt{2}) + \frac34 \ln 2 / (2N+1)$, the latter term being a contribution from $f_0$ of (\ref{f0dimer}).  There are additional corrections that decay exponentially with both $N$ and $M$.  For example, for $N = 5$, $a(5) = -0.393426758471593423337857$, but actually $(2 M + 1) (f(5,M)-f^\mathrm{strip}(5)) \to -0.3934267586439834913176$ as $M \to \infty$, the difference being just$-1.72390067979787\cdot10^{-10}$.  As $N$ increases, the difference between the observed coefficient to the $1/(2M+1)$ and $a(n)$ decays exponentially.   Likewise, the additional corrections to Eq.\ (\ref{largeM}) with respect to $M$ also decay exponentially.  Plotting $f(N,M)$ vs.\ $1/(2M+1)$, for example for $M=10,000, 100,000,$ and 1,000,000, yields a nearly perfect straight line with slope $-0.393426758638$ (showing a very small correction for small $N$ to the value for $a(5)$ given above) and intercept of $0.2525855505199970$, which exactly confirms the prediction of $f^\mathrm{strip}(5)$ from Eq.\ (\ref{exactstrip}).   For large $N$ and $M$ with $M \gg N$, Eq.\ (\ref{largeM}) exhibits essentially the entire correction between $f(N,M)$ and $f^\mathrm{strip}(N)$.


\begin{table}[ht]
	\caption{The free energy $f^\mathrm{strip}(N)$ for an infinite strip of width $2N$, both exact using Eq.\ (\ref{exactstrip}) and the asymptotic series, Eq.\ (\ref{expansion1}), cut off at the term $p = p_\mathrm{max}$ where the contribution of the last term to the sum is at a minimum.   This cutoff corresponds almost exactly to point where the difference between the exact and asymptotic series is at a minimum; that difference is shown in the third column.   It can be seen that the number of useful digits in the asymptotic series (the error), and also the number of terms used in the asymptotic series, grows roughly linearly with $N$.}
	\label{table:strip}
	\begin{center}
		\begin{tabular}{| c | c | c | c |}
			\hline
			$N$ & $f^\mathrm{strip, exact}(N)$, Eq.\ (\ref{exactstrip}) & $f^\mathrm{strip, exact}(N) - f^\mathrm{strip,asympt}_{p_\mathrm{max}}(N)$, Eq.\ (\ref{expansion1}) & $p_\mathrm{max}$ \\\hline
			1 & 0.1604039416865344824992529711414561410451 & $-5.132227939\times10^{-6} $& 2 \\\hline
			2 & 0.2087985670180315478881804103188546040574 & $-6.111553441\times10^{-7}$ & 4  \\\hline
			3 & 0.2313105474413464893025375385071050814009 & $-1.775947573\times10^{-8}$ & 6 \\\hline
			4 & 0.244223653343733699288287454551312000794 &$-4.717421345\times10^{-10}$&  8\\\hline
			5 & 0.2525855505199976248740884126691071592619& $-1.278690965\times10^{-11}$ &10 \\\hline
			6 & 0.2584391557216788782944540541059248046135 & $-5.519215239\times10^{-14}$  & 11  \\\hline
			7 & 0.2627650673950520421902381856753860728828&$-2.615778603\times10^{-15}$  & 13\\\hline
			8 &0.2660920022673037714519038101971489895370& $-9.221773685\times10^{-17}$& 15\\\hline
			9 & 0.2687300478449198125301383082353948301983& $-2.943374437\times10^{-18} $& 17 \\\hline
			10 & 0.2708730331629703759051898185982918361479 &$ -8.976651599\times10^{-20}$ & 19 \\              \hline
			20 & 0.2808903424969997495135060643558988391223 & $-5.205122376\times10^{-36} $& 36 \\  \hline
			50 & 0.2872105012211973964642182161192474498935 & $-1.223683572\times10^{-82}$ & 89  \\  \hline 
			100 & 0.2893716724787264395315835716094691245145 & $-6.780514699\times10^{-160}$ & 177 \\                  \hline
		\end{tabular}
	\end{center}
\end{table}

\section{The ratio of the coefficients $f_p(\rho)$ in the free energy expansion}
\label{ratiofreeenergy}
Let us consider the ratio of the coefficients $f_p$ in the free energy expansion for strip $f_p^\mathrm{strip}$ and square $f_p^\mathrm{sq}=f_p(1)$ geometry
$$
r_p=\frac{f_p^\mathrm{strip}}{f_p^\mathrm{sq}}
$$
One can clearly see from the Table \ref{tabular1} that the ratio  tends to 1/2 as $p$ tends to $\infty$
\begin{equation}
\lim_{p \to \infty}r_p=\frac{1}{2} \label{rprho1}
\end{equation}
Let us consider ratio of coefficients $f_p$ in free energy expansion for rectangular geometry (for arbitrary $\rho > 1$) and square geometry ($\rho=1$) multiplied by  $\rho^{-p-1}$
\begin{equation}
r_p(\rho)=\frac{f_p(\rho)\rho^{-p-1}}{f_p^\mathrm{sq}} \hspace{1cm} \mbox{for} \quad \rho > 1 \label{rprb1}
\end{equation}
Again one can see from Table \ref{tabular1} that for $\rho=2$ the ratio $r_p(2)$ also tends to 1/2 as $p$ tends to $\infty$. Furthermore, 
we find that for arbitrary $\rho > 1$, $r_p(\rho)$ is equal to $\frac{1}{2}$ in the limit ${p \to \infty}$, namely
\begin{equation}
\lim_{p \to \infty} r_p(\rho)=\frac{1}{2} \hspace{1cm} \mbox{for} \quad \rho > 1. \label{rprho2}
\end{equation}
It is easy to see from Eqs.\ (\ref{Jacobifp}) and (\ref{rprb1}) that $r_p(\rho)$ transforms under Jacobi transformation  $\rho \to 1/\rho$ as
\begin{equation}
r_p\left(\frac{1}{\rho}\right)=\bar r_p(\rho)\label{Jacobirp}
\end{equation}
where the ratio $\bar r_p(\rho)$ is given by
$$
\bar r_p(\rho)=\frac{f_p(\rho)\rho^{p+1}}{f_p^\mathrm{sq}}. 
$$ 
Thus one can obtain from Eq.\  (\ref{Jacobirp}) that for $\rho < 1$ the ratio $\bar r_p(\rho)$ tends to 1/2 as $p$ tends to infinity
\begin{equation}
\lim_{p \to \infty} \bar r_p(\rho)=\frac{1}{2} \hspace{1cm} \mbox{for} \quad  \rho < 1. \label{rprho3}
\end{equation}
Thus if the shape $\rho$ of the rectangular $2M \times 2N$ lattice is varied, the ratios $r_p(\rho)$ (for $\rho>1$) and $\bar r_p(\rho)$ (for $\rho<1$) tend to 1/2 as $p$ tends to infinity.  

\begin{figure}[!ht]
	\includegraphics[width=150mm]{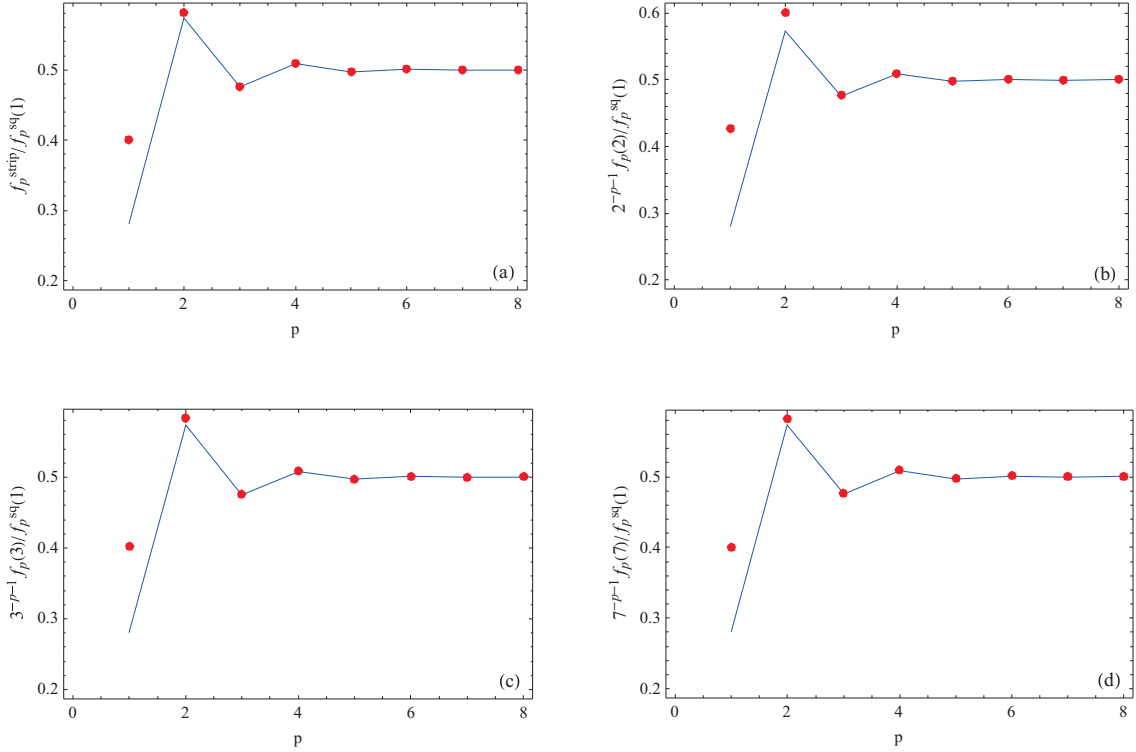}
	\caption{(Color online) (a) The ratio $f_p^{\mathrm{strip}}/f_p^{\mathrm{sq}}(1)$ as a function of order $p$ for the square ($\rho = 1$). Dots represent our exact results. The solid line is given by $(-1)^p a \, b^p + 1/2$, with $a=0.6561$ and $b=0.3348$ for all four plots, and represents the exponential decay to 1/2.  (b) The ratio $\rho^{-p-1}f_p(\rho)/f_p^{\mathrm{sq}}(1)$ as function of $p$ for $\rho = 2$, (c) for $\rho = 3$, and (d) for $\rho = 7$.}
	\label{fig3}
\end{figure}

In Fig.\ \ref{fig3} we plot the ratio $\rho^{-p-1}f_p(\rho)/f_p^{\mathrm{sq}}(1)$ as a function of $p$ for (a) $\rho =1$, (b) $\rho = 2$, (c) $\rho = 3$ and (d) $\rho = 7$.

\begin{figure}[!ht]
	\includegraphics[width=120mm]{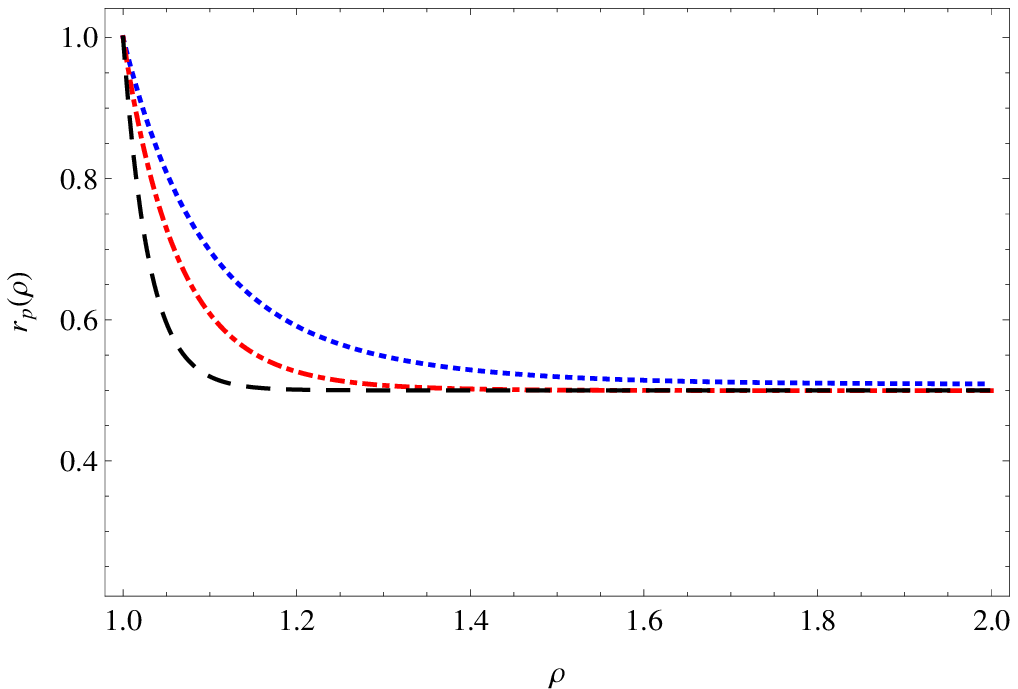}
	\caption{Plot of the ratio (\ref{rprb1}) as a function of $\rho$ for various $p$ (dotted line for $p=5$, dotted-dashed line for $p=7$, and dashed line for $p=16$),  showing that $r(\rho)$ becomes a step function at $\rho = 1$ as $p \to \infty$.}
	\label{fig:step}
\end{figure}

Note that Eq.\ (\ref{rprho2}) cannot apply when $\rho = 1$, because then the rectangle is a square and $ r_p(\rho=1)$ should be 1.  Indeed, we find that as $p$ becomes large, the behavior of $ r_p(\rho)$ as a function of $\rho$ becomes a step function at $\rho = 1$, as shown in Fig.\ \ref{fig:step}.

The results given by Eqs.\ (\ref{rprho1}), (\ref{rprho2}) and (\ref{rprho3}) are very surprising results, and we do not have any physical explanation why those ratios should be finite and equal $1/2$.  To get more insight on these effects we plan to consider the ratio $r_p(\rho)$ in other models, such as the spanning tree model under different boundary conditions and the Ising model with Brascamp-Kunz boundary conditions, in the near future. 

\begin{table}[ht]
	\caption{Ratios of asymptotic coefficients $f_p$ for the strip to the square and the rectangle of aspect ratio 2 to the square, as a function of $p$.}
	\label{tabular1}
	\begin{center}
		\begin{tabular}{| c | c | c |}
			\hline
			$p$ & $r_p=\frac{f_p^\mathrm{strip}}{f_p^\mathrm{sq}(1)}$ & $r_p=\frac{2^{-p-1} f_p(2)}{f_p^\mathrm{sq}(1)}$ \\\hline
			0 & 0.188848339243... & 0.565194162895... \\\hline
			1 & 0.400706280144... & 0.426316738242... \\\hline
			2 & 0.582023532257... & 0.601336607490... \\\hline
			3 & 0.475899439722... & 0.476675704408... \\\hline
			4 & 0.508793966746... & 0.509148251059... \\\hline
			5 & 0.496962636744... & 0.497116857889... \\\hline
			6 & 0.501071217692... & 0.501105333933... \\\hline
			7 & 0.499636849243... & 0.499643260737... \\\hline
			8 & 0.500121853753... & 0.500123707369... \\\hline
			9 & 0.499959350873... & 0.499959866780... \\\hline
			10 & 0.500013502309... & 0.500013621630... \\\hline
			11 & 0.499995541158... & 0.499995569751... \\\hline
			12 & 0.500001465697... & 0.500001473193... \\\hline
			13 & 0.499999519812... & 0.499999521710... \\\hline
			14 & 0.500000156880... & 0.500000157343... \\\hline
			15 & 0.499999948878... & 0.499999948993... \\\hline
			16 & 0.500000016621... & 0.500000016651... \\\hline
			17 & 0.499999994606... & 0.499999994614... \\\hline
			18 & 0.500000001747... & 0.500000001749... \\\hline
			19 & 0.499999999435... & 0.499999999435... \\\hline
			20 & 0.500000000183... & 0.500000000183... \\\hline
			21 & 0.499999999941... & 0.499999999941... \\\hline
			22 & 0.500000000019... & 0.500000000019... \\\hline
		\end{tabular}
	\end{center}
\end{table}
 
\section{Conclusion}
\label{Conclusion}
We have analyzed the partition function of the dimer model on a rectangular $2M \times 2N$ lattice with free boundary conditions. We have obtained that the corner free energy for the dimer model on a rectangular $2M \times 2N$ lattice with free boundary conditions is equal to zero, due to the strong nonlocality of the dimer model,  and explain that result within the framework of conformal field theory, in which the central charge of the considering models is $c=-2$. 
Using the results of Ref.\ \cite{Izmailian2003} we have calculated the exact expressions for the correction terms $f_p(\rho)$ in the asymptotic expansion of the free energy in the dimer model from $p=0$ up to $p=22$.  We also found the general asymptotic formula as well as a simple exact expression for the strip.   We find that ratio of coefficients $f_p(\rho)$ in the free energy expansion for the strip  ($f_p^\mathrm{strip}$)  and square ($f_p^\mathrm{sq}$)  geometry	$r_p={f_p^\mathrm{strip}}/{f_p^\mathrm{sq}}$, in the limit of large $p$, tends to $1/2$, and predict that ratio is exactly equal to 1/2 in the limit of $p \to \infty$. Furthermore, we predict that the ratio of coefficients $f_p(\rho)$ in free energy expansion for rectangular ($f_p(\rho)$) for $\rho > 1$  and square ($f_p^\mathrm{sq}$) geometry multiplied by $\rho^{-p-1}$, that is	$r_p={\rho^{-p-1} f_p(\rho) }/{f_p^\mathrm{sq}}$, is also equal to $1/2$  in the limit of $p \to \infty$, and using the Jacobi transformation we find that the ratio $r_p=\rho^{p+1} {f_p^\mathrm{sq}}/{f_p(\rho)}$ is also equal to  $1/2$ for $\rho < 1$ in the limit of $p \to \infty$. These are  very surprising simple results, and to gain some insight on them we plan to study the ratio $r_p(\rho)$ in other models, such as the spanning tree model under different boundary conditions and the Ising model with Brascamp-Kunz boundary conditions.
\section{Acknowledgments}
\label{Acknowledgment}

This work was supported by a grant of the Science Committee of the Ministry of Education and Science of the Republic of Armenia under contract 18T-1C113 and by the JINR program ``Ter-Antonyan-Smorodinsky."  R. Z. would like to thank Walter Trump for discussions related to the numerical evaluation of the number of dimer configurations.

\appendix

\section{Expressions for $\Lambda_{2p}$}
\label{Lambda}
The differential operators $\Lambda_{2p}$ that appear in Eq.\  (\ref{ExpansZab})
can be expressed via coefficients $z_{2p}$ of the Taylor series expansion
of the lattice dispersion relation $\omega_{z}(k)$ of Eq.\ (\ref{omega}) about $k = 0$, taking $z=1$ here because we will eventually take $z = 1$ when evaluating (\ref{fpoocc}).  We have, for $z=1$:
\begin{equation}
\omega _z(k)=k\left(1 +\sum _{p=1}^{\infty}
\frac{z _{2p}}{(2p)!}k^{2p}\right)
\end{equation}
with $z_2 = -2/3$, $z_4 = 4$, $z_6 = -632/7$, $z_8 = 39440/9$, $z_{10} = -4087712/11$, $z_{12} = 48787520$, $z_{14} = -137605112192/15$, $z_{16} = 2339775197440$, $z_{18} = -14775064298435072/19$, $z_{20} = 6857795892626969600/21$, etc. These $z_{2p}$ can be found simply from Eq.\ (\ref{omega}) using Mathematica. Then we have
\begin{eqnarray}
{\Lambda}_{2}&=&-\frac{2}{3},
\nonumber\\
{\Lambda}_{4}&=&4+\frac{4}{3}\,\frac{\partial}{\partial z},
\nonumber\\
{\Lambda}_{6}&=&-\frac{632}{7}-40\,\frac{\partial}{\partial z}
-\frac{40}{9}\,\frac{\partial^2}{\partial z^2},
\nonumber\\
&\vdots&
 \nonumber\\
{\Lambda}_{p}&=&\sum_{r=1}^{p}\sum
\left(\frac{ z_{p_1}}{p_1!}\right)^{k_1}\ldots
\left(\frac{ z_{p_r}}{p_r!}\right)^{k_r}\frac{p!}{k_1!\ldots
	k_r!}\;\frac{\partial^k}{\partial z^k}.
\label{L2p}
\end{eqnarray}
Here summation is over all positive numbers $\{k_1\ldots k_r\}$
and different positive numbers $\{p_1,\ldots,p_r\}$ such that $p_1
k_1+\ldots+ p_r k_r=p$ and $k_1+\ldots+k_r-1 = k$.  
In Supplementary Materials we have listed expressions for the ${\Lambda}_{2p}$ up to $p = 22$. 
\section{Jacobi transformation}
\label{Jacobi}
Let us consider the behavior of the theta functions, the Dedekind $\eta$ function, and the Kronecker functions $K_{2p}^{\frac{1}{2},\frac{1}{2}}$ under the Jacobi
transformation (\ref{r1r}).
The result for the theta functions and the Dedekind $\eta$
function  is given by \cite{Korn}:
\begin{eqnarray}
\theta_3\left(\frac{1}{\rho}\right)&=&\sqrt{\rho}\,\theta_3(\rho) \label{theta3}\\
\theta_2\left(\frac{1}{\rho}\right)&=&\sqrt{\rho}\,\theta_4(\rho) \label{theta2}\\
\theta_4\left(\frac{1}{\rho}\right)&=&\sqrt{\rho}\,\theta_2(\rho) \label{theta4}\\
\eta\left(\frac{1}{\rho}\right)&=&\sqrt{\rho}\,\eta(\rho) \label{eta1}
\end{eqnarray}

The result for the Kronecker functions $K_{2p}^{\frac{1}{2},\frac{1}{2}}$ can be obtained from
the relation between the coefficients in the Laurent expansion
of the Weierstrass function and Kronecker functions
(see Appendix (\ref{KroneckerToTheta})) and is given by
\begin{eqnarray}
K_{2p}^{\frac{1}{2},\frac{1}{2}}\left(\frac{1}{\rho}\right)&=&\rho^{2p}K_{2p}^{\frac{1}{2},\frac{1}{2}}(\rho) 
\label{Kronecker}
\end{eqnarray}

\section{Reduction of Kronecker's Double Series to Theta Functions}
\label{KroneckerToTheta}
The Kronecker functions ${\rm K}_{2p}^{0,0}(\tau)$ are related
directly to the coefficients $a_p(\tau)$
\begin{equation}
{\rm K}^{0,0}_{2p}(\tau)=
-\frac{(2p)!}{(-4\pi^2)^p}\frac{a_p(\tau)}{(2p-1)}\label{Kabat}
\end{equation}
where $a_p(\tau)$ are the coefficients in the Laurent expansion of the Weierstrass function $\wp(z)$ with two periods $\omega_1 =1$ and $\omega_2=\tau$
\begin{eqnarray}
\wp(z)&=&\frac{1}{z^2}+\sum_{(n,m)\neq(0,0)}\left[\frac{1}{(z-n-\tau
	m)^2}-\frac{1}{(n+\tau m)^2}\right]\nonumber\\
&=&\frac{1}{z^2}+\sum_{p=2}^{\infty}a_{p}(\tau) z^{2p-2}\nonumber
\end{eqnarray}

The coefficients $a_p(\tau)$ can all be written
in terms of the elliptic theta functions with the help of the recursion relation (see Ref.\ \cite{Korn} page 749)
$$a_p=\frac{3}{(p-3)(2p+1)}~
(a_{2}a_{p-2}+a_{3}a_{p-3}+\ldots+a_{p-2}a_{2})$$
where first terms of the sequence are
\begin{eqnarray}
a_2&=&\frac{\pi^4}{15}(\theta_2^8+
\theta_2^4\theta_4^4+\theta_4^8)\nonumber\\
a_3&=&\frac{\pi^6}{189}\left[2(\theta_4^{12}-\theta_2^{12})+
3(\theta_4^4-\theta_2^4)\theta_2^4\theta_4^4\right]\nonumber\\
a_4&=&\frac{1}{3}a_2^2\nonumber\\
a_5&=&\frac{3}{11}a_2a_3\nonumber\\
&\vdots&\nonumber
\end{eqnarray}
The coefficients $a_p$ are the coefficients in the Laurent expansion of
the Weierstrass function. The coefficients $a_2$ and $a_3$ are
related to coefficients $g_2$ and $g_3$ (see for example 
\cite{Korn} p.\ 749 and 767.)

The Kronecker function ${\rm K}_{2p}^{\frac{1}{2},\frac{1}{2}}(\tau)$ can in turn be related to the function ${\rm K}_{2p}^{0,0}(\tau)$ by means of
simple resummation of Kronecker's double series
\begin{eqnarray}
{\rm K}_{2p}^{\frac{1}{2},\frac{1}{2}}(\tau)&=& (1+2^{2-2p})\,{\rm
	K}^{0,0}_{2p}(\tau)-2^{1-2p}\,{\rm K}^{0,0}_{2p}(\tau/2)-2\,{\rm
	K}^{0,0}_{2p}(2\tau)\nonumber
\end{eqnarray}
Thus, Kronecker functions  ${\rm K}_{2p}^{\frac{1}{2},\frac{1}{2}}(\tau)$ can all be
expressed in terms of the elliptic $\theta$-functions only. For
practical calculations the following identities are also helpful

\begin{center}
	\begin{tabular}{rclcrcl}
		$2\theta_2^2(2\tau)$&$=$&$\theta_3^2-\theta_4^2$&~~~~~~~~~~~~~
		&$\theta_2^2(\tau/2)$&$=$&$2\theta_2\theta_3$\\[0.15cm]
		$2\theta_3^2(2\tau)$&$=$&$\theta_3^2+\theta_4^2$&
		&$\theta_3^2(\tau/2)$&$=$&$\theta_2^2+\theta_3^2$\\[0.15cm]
		$2\theta_4^2(2\tau)$&$=$&$2\theta_3\theta_4$&
		&$\theta_4^2(\tau/2)$&$=$&$\theta_3^2-\theta_2^2$
	\end{tabular}
\end{center}
From the general formulas above we can easily write down the
Kronecker functions  ${\rm K}_{2p+2}^{\frac{1}{2},\frac{1}{2}}(\tau)$  for all value of $p$. In particular in Supplementary Materials we have listed expressions for the Kronecker functions  ${\rm K}_{2p+2}^{\frac{1}{2},\frac{1}{2}}(\tau)$ up to $p = 11$.

\section{Expressions for $f_p(\rho)$ from $p=0$ up to $p=5$.}
These give the explicit asymptotic coefficients $f_p(\rho)$  that enter in Eq.\ (\ref{expansion}) for an arbitrary aspect ratio $\rho$, for order $p = 0$ to $p = 5$.
\begin{eqnarray}
f_{0}&=&\frac{3}{4}\ln 2 + \frac{1}{2} \ln \frac{\theta _3}{\eta } \cr
f_{1}&=& \frac{\pi ^3 \rho ^2}{12} \left(\frac{11}{120} \theta _2^4 \theta
_4^4+\frac{7}{240} \left(\theta _2^8+\theta
_4^8\right)\right) \cr
f_{2}&=&
\frac{\pi ^6\rho ^4}{48384}\theta _4^4 \left(-5 \theta _2^{12}+5 \theta
_4^4 \theta _2^8+41 \theta _4^8 \theta _2^4+31 \theta
_4^{12}\right)  \cr
&+&\frac{\pi ^5\rho ^3}{48384} \left(31 \theta _2^{12}+15 \theta _4^4 \theta _2^8-15
\theta _4^8 \theta _2^4-31 \theta _4^{12}\right) \left(2\rho
E \sqrt{\theta _2^4+\theta _4^4}
-1\right)
\cr
f_{3}&=&\frac{\pi ^9 \rho ^6}{829440}\theta
_4^4 \left(\theta _2^4+\theta _4^4\right){}^2 \left(5 \theta
_2^{12}+119 \theta _4^4 \theta _2^8+431 \theta _4^8 \theta
_2^4+381 \theta _4^{12}\right)\cr
&+&\frac{\pi ^8 \rho ^5}{138240}\theta _4^4 \left(71
\theta _2^{12}+93 \theta _4^4 \theta _2^8+213 \theta _4^8
\theta _2^4+127 \theta _4^{12}\right)  \left[\theta _2^4+\theta _4^4-2 \rho
E \left(\theta _2^4+\theta _4^4\right){}^{3/2}\right]\cr
&+&\frac{\pi^7\rho^4}{19353600} \left(127 \theta _2^{16}+284 \theta _4^4 \theta
_2^{12}+186 \theta _4^8 \theta _2^8+284 \theta _4^{12} \theta
_2^4+127 \theta _4^{16}\right) \cr
&\times& \left[280\rho ^2 E^2
\left(\theta _2^4+\theta _4^4\right) -280\rho E
\sqrt{\theta _2^4+\theta _4^4}
+79\right]\cr
f_{4}&=&\frac{\pi ^{12}\rho ^8\theta
	_4^4 \left(\theta _2^4+\theta _4^4\right){}^2}{7962624}  \left(\theta
_2^{20}+10 \theta _4^4 \theta _2^{16}+1470 \theta _4^8 \theta
_2^{12}+6744 \theta _4^{12} \theta _2^8+10137 \theta _4^{16}
\theta _2^4+5110 \theta _4^{20}\right)\cr
&+&\frac{\pi ^{11}\rho ^7 \theta _4^4}{1327104} \left(3 \theta _2^{16}-620 \theta
_4^4 \theta _2^{12}+1626 \theta _4^8 \theta _2^8+3780 \theta
_4^{12} \theta _2^4+2555 \theta _4^{16}\right)
\left[\left(\theta _2^4+\theta _4^4\right){}^2-2\rho E
\left(\theta _2^4+\theta _4^4\right){}^{5/2}
\right]\cr
&+&\frac{\pi ^{10}\rho ^6 \theta _4^4 \left(\theta
	_2^4+\theta _4^4\right)}{510935040} \left(-1261 \theta _2^{16}-2012
\theta _4^4 \theta _2^{12}+3018 \theta _4^8 \theta _2^8+5044
\theta _4^{12} \theta _2^4+2555 \theta _4^{16}\right) \cr
&\times&\left[421-1540\rho E  \left(\sqrt{\theta _2^4+\theta
	_4^4}-\rho E \left(\theta _2^4+\theta _4^4\right)
\right)\right]\cr
&+&\frac{\pi ^9\rho ^5}{306561024} \left(511 \theta
_2^{20}+1261 \theta _4^4 \theta _2^{16}+1006 \theta _4^8
\theta _2^{12}-1006 \theta _4^{12} \theta _2^8-1261 \theta
_4^{16} \theta _2^4-511 \theta _4^{20}\right) \cr
&\times& \left[-493+2\rho
E \left(770\rho E \left(\theta _2^4+\theta
_4^4\right)   \left(2 \rho E \sqrt{\theta _2^4+\theta
	_4^4}  -3\right)+1263 \sqrt{\theta _2^4+\theta
	_4^4}\right)\right]\cr
f_{5}&=&\frac{\pi ^{15}  \rho ^{10} \theta _4^4
	\left(\theta _2^4+\theta _4^4\right){}^2}{955514880} \left(2 \theta _2^{28}+33 \theta _4^4 \theta
_2^{24}+112308 \theta _4^8 \theta _2^{20}+817659 \theta _4^{12} \theta _2^{16}+2737458
\theta _4^{16} \theta _2^{12}\right.\cr
&+&\left.4867063 \theta _4^{20} \theta _2^8+4245368 \theta
_4^{24} \theta _2^4+1414477 \theta _4^{28}\right)\cr
&+&\frac{\pi ^{14}\rho ^9 \theta _4^4}{238878720} \left(15 \theta
_2^{24}+180 \theta _4^4 \theta _2^{20}+457799 \theta _4^8 \theta _2^{16}+1505408
\theta _4^{12} \theta _2^{12}+3187197 \theta _4^{16} \theta _2^8+3538540 \theta
_4^{20} \theta _2^4\right.\cr
&+&\left.1414477 \theta _4^{24}\right) \left(\left(\theta
_2^4+\theta _4^4\right){}^2-2 \rho E \left(\theta _2^4+\theta _4^4\right){}^{5/2}
\right)\cr
&+&\frac{\pi ^{13} \rho ^8 \theta _4^4 \left(\theta
	_2^4+\theta _4^4\right){}^2}{39016857600} \left(105 \theta _2^{20}+354433 \theta _4^4 \theta
_2^{16}+1097690 \theta _4^8 \theta _2^{12}+1860642 \theta _4^{12} \theta _2^8+2831789
\theta _4^{16} \theta _2^4\right.\cr
&+&\left.1414477 \theta _4^{20}\right) \left[263-980
\rho E  \left(\sqrt{\theta _2^4+\theta _4^4}- \rho E \left(\theta
_2^4+\theta _4^4\right) \right)\right]\cr
&-&\frac{263 \pi ^{12}\rho ^8 E \theta _4^4 \sqrt{\theta
		_2^4+\theta _4^4}}{9754214400} \left(-707921 \theta _2^{24}+2214130 \theta _4^4 \theta
_2^{20}+3899371 \theta _4^8 \theta _2^{16}+5405580 \theta _4^{12} \theta
_2^{12}\right.\cr
&+&\left.6552023 \theta _4^{16} \theta _2^8+4954082 \theta _4^{20} \theta _2^4+1414477
\theta _4^{24}\right)\cr
&-&\frac{\pi ^{12}\rho ^7 \theta _4^4 \left(\theta _2^4+\theta _4^4\right)}{58525286400} \left(707921 \theta
_2^{20}+1506209 \theta _4^4 \theta _2^{16}+2393162 \theta _4^8 \theta _2^{12}+3012418
\theta _4^{12} \theta _2^8\right.\cr
&+&\left.3539605 \theta _4^{16} \theta _2^4+1414477 \theta
_4^{20}\right)  \left[-299 + 980\rho^2 E^2 \left(\theta _2^4+\theta _4^4\right)
\left(2 \rho E \sqrt{\theta _2^4+\theta _4^4}
-3\right)\right]\cr
&+&\frac{\pi ^{11}\rho ^6}{83691159552000}
\left(1414477 \theta _2^{24}+4247526 \theta _4^4 \theta _2^{20}+4518627 \theta _4^8
\theta _2^{16}+4786324 \theta _4^{12} \theta _2^{12}+4518627 \theta _4^{16} \theta_2^8\right.\cr
&+&\left.4247526 \theta _4^{20} \theta _2^4+1414477 \theta _4^{24}\right)
\left[127741-2860 \rho E \left(299 \sqrt{\theta _2^4+\theta _4^4}\right.\right.\cr
&-&\left.\left.\rho E
\left(\theta _2^4+\theta _4^4\right)   \left(789+490\rho E   \left(\rho E
\left(\theta _2^4+\theta _4^4\right) -2 \sqrt{\theta _2^4+\theta
	_4^4}\right)\right)\right)\right]
\cr
\nonumber
\end{eqnarray}

\vfill\eject

\begin{center}
\bf{SUPPLEMENTARY MATERIALS}
\end{center}
\section{Expressions for $\Lambda_{2p}$  from $p=1$ up to $p=22$}
\label{LambdaSupp}
The operators $\Lambda_{2p}$  are discussed in Appendix \ref{Lambda} of the paper.
\begin{eqnarray}
{\Lambda}_{2}&=&-\frac{2}{3},
\nonumber\\
{\Lambda}_{4}&=&4+\frac{4}{3}\,\frac{\partial}{\partial z},
\nonumber\\
{\Lambda}_{6}&=&-\frac{632}{7}-40\,\frac{\partial}{\partial z}
-\frac{40}{9}\,\frac{\partial^2}{\partial z^2},
\nonumber\\
{\Lambda}_{8}&=&\frac{39440}{9}+\frac{6736}{3}\,\frac{\partial}{\partial z}
+\frac{1120}{3}\,\frac{\partial^2}{\partial z^2}+\frac{560}{27}\,\frac{\partial^3}{\partial z^3},
\nonumber\\
{\Lambda}_{10}&=&-\frac{4087712}{11}-\frac{621920}{3}\,\frac{\partial}{\partial z}
-42080\,\frac{\partial^2}{\partial z^2}-\frac{11200}{3}\,\frac{\partial^3}{\partial z^3}-\frac{1120}{9}\,\frac{\partial^4}{\partial z^4},
\nonumber\\
{\Lambda}_{12}&=&48787520+\frac{201555520 }{7}\,\frac{\partial}{\partial z}
+\frac{19796480}{3}\,\frac{\partial^2}{\partial z^2}+\frac{2221120}{3}\,\frac{\partial^3}{\partial z^3}+\frac{123200}{3}\,\frac{\partial^4}{\partial z^4}+\frac{24640}{27}\,\frac{\partial^5}{\partial z^5},
\nonumber\\
{\Lambda}_{14} &=& -\frac{137605112192}{15} - 5635846528 \,\frac{\partial}{\partial z} - 1402670464 \,\frac{\partial^2}{\partial z^2} -
\frac{4917552640}{27}\,\frac{\partial^3}{\partial z^3}\nonumber\\
&-& \frac{117877760}{9}\,\frac{\partial^4}{\partial z^4} - \frac{4484480}{9} \,\frac{\partial^5}{\partial z^5}- \frac{640640}{
	81}\,\frac{\partial^6}{\partial z^6}\nonumber\\
{\Lambda}_{16} &=& 2339775197440 + \frac{13331890128640 }{9}\,\frac{\partial}{\partial z} + \frac{1174430067200}{
	3}\,\frac{\partial^2}{\partial z^2}+\frac{506157733120}{9}\,\frac{\partial^3}{\partial z^3}\nonumber\\ &+&\frac{128678950400}{27}\,\frac{\partial^4}{\partial z^4} + \frac{2155112960}{
	9}\,\frac{\partial^5}{\partial z^5} + \frac{179379200 }{27}\,\frac{\partial^6}{\partial z^6} + \frac{6406400 }{81}\,\frac{\partial^7}{\partial z^7}\nonumber\\
{\Lambda}_{18} &=&-\frac{14775064298435072}{19}-503972805294592 \,\frac{\partial}{\partial z}-\frac{2927802536177152}{21}\,\frac{\partial^2}{\partial z^2}\nonumber\\
&-&\frac{64743412065280}{3}\,\frac{\partial^3}{\partial z^3}-\frac{6155889159680}{3}\,\frac{\partial^4}{\partial z^4}-\frac{1108691584000}{9}\,\frac{\partial^5}{\partial z^5}-\frac{41211089920}{9}\,\frac{\partial^6}{\partial z^6}\nonumber\\
&-&\frac{871270400}{9}\,\frac{\partial^7}{\partial z^7}-\frac{217817600}{243}\,\frac{\partial^8}{\partial z^8}\nonumber\\
{\Lambda}_{20} &=& \frac{6857795892626969600}{21}+\frac{16604069879200435200 }{77}\,\frac{\partial}{\partial z}+\frac{432908398929408000 }{7}\,\frac{\partial^2}{\partial z^2}\nonumber\\
&+&\frac{212949248427904000}{21}\,\frac{\partial^3}{\partial z^3}+1050679396582400\,\frac{\partial^4}{\partial z^4}+71554442818560\,\frac{\partial^5}{\partial z^5}\nonumber\\
&+&\frac{260264151347200}{81}\,\frac{\partial^6}{\partial z^6}+\frac{2485485516800}{27}\,\frac{\partial^7}{\partial z^7}+\frac{41385344000}{27}\,\frac{\partial^8}{\partial z^8}+\frac{8277068800}{729}\,\frac{\partial^9}{\partial z^9} \nonumber
\end{eqnarray}
\begin{eqnarray}
{\Lambda}_{22} &=&
-\frac{3889298341511511652352}{23}-\frac{340154164075949516800}{3}\,\frac{\partial}{\partial z}-33481448867175389184 \,\frac{\partial^2}{\partial z^2}\nonumber\\
&-&5747611413630648320 \,\frac{\partial^3}{\partial z^3}
-\frac{17192013187992924160}{27}\,\frac{\partial^4}{\partial z^4}-\frac{429147526308300800}{9}\,\frac{\partial^5}{\partial z^5}\nonumber\\
&-&\frac{66167334948915200}{27}\,\frac{\partial^6}{\partial z^6}-\frac{6932012011724800}{81}\,\frac{\partial^7}{\partial z^7}-\frac{52625603430400}{27}\,\frac{\partial^8}{\partial z^8}\nonumber\\
&-&\frac{6373342976000}{243}\,\frac{\partial^9}{\partial z^9}-\frac{115878963200 \text{d10}}{729}\,\frac{\partial^{10}}{\partial z^{10}}
\nonumber\\
{\Lambda}_{24} &=&\frac {529272525177347648180224} {5} +
\frac {359448887615106330472448} {5}\frac {\partial} {\partial z}\cr\cr
&+&\frac {587161740196038302556160} {27}\frac {\partial^2} {\partial z^2} +
\frac {244168714426028141342720} {63}\frac {\partial^3} {\partial z^3} \cr\cr
&+& \frac {4077658907784735539200} {9}\frac {\partial^4} {\partial z^4} +
\frac {986927466800024240128} {27}\frac {\partial^5} {\partial z^5} \cr\cr
&+& \frac {56188334282254254080} {27}\frac {\partial^6} {\partial z^6} +
\frac {2262298979469168640} {27}\frac {\partial^7} {\partial z^7} \cr\cr
&+&\frac {189817455949721600} {81}\frac {\partial^8} {\partial z^8} +
\frac {10562632361881600} {243}\frac {\partial^9} {\partial z^9} \cr\cr
&+&\frac {117269510758400} {243} \frac {\partial^{10}} {\partial z^{10}} +
\frac {5330432307200} {2187}\frac {\partial^{11}} {\partial z^{11}}
\cr\cr
{\Lambda} _ {26} & = & - \frac {2127690488032789501903020032} {27}
-\frac {3408308674717223537465221120 } {63}\frac {\partial} {\partial z}\cr\cr
& - &\frac {1052142278426714983005921280 } {63}\frac {\partial^{2}} {\partial z^{2}}
-\frac {1741109452063314562547302400 } {567}\frac {\partial^{3}} {\partial z^{3}}\cr\cr
& - &\frac {23637506964029074546892800 } {63}\frac {\partial^{4}}{\partial z^{4}}
-\frac {289342086671401713664000 } {9}\frac {\partial^{5}} {\partial z^{5}}\cr\cr
& - &\frac {160730205258885177344000 } {81}\frac {\partial^{6}} {\partial z^{6}}
-\frac {2404423382287646720000 } {27}\frac {\partial^{7}} {\partial z^{7}}\cr\cr
& - &\frac {78013037272721408000 } {27}\frac {\partial^{8}} {\partial z^{8}}
-\frac {144844179754193920000 } {2187}\frac {\partial^{9}} {\partial z^{9}}\cr\cr
& - &\frac {743739990017024000 } {729}\frac {\partial^{10}} {\partial z^{10}}
-\frac {6929561999360000 } {729}\frac {\partial^{11}} {\partial z^{11}}
-\frac {266521615360000 } {6561}\frac {\partial^{12}} {\partial z^{12}}
\cr\cr
{\Lambda} _ {28} & = & 68811534804911910033805557760
+\frac {143041939473292631263408504832 } {3}\frac {\partial}{\partial z}\cr\cr
& + & 14975776132932596018445025280\frac {\partial^{2}} {\partial z^{2}}
+\frac {8481865264365212531060654080 } {3}\frac {\partial^{3}}
{\partial z^{3}}\cr\cr
& + &\frac {3226421315052535209146368000 } {9}\frac {\partial^{4}}
{\partial z^{4}}
+32286528854481337813811200\frac {\partial^{5}} {\partial z^{5}}\cr\cr
& + &\frac {6384105020273642114252800 } {3}\frac {\partial^{6}} {\partial z^{6}}
+\frac {2810396463616900289331200 } {27}\frac {\partial^{7}} {\partial z^{7}}\cr\cr
& + &\frac {34057859162147172352000 } {9}\frac {\partial^{8}} {\partial z^{8}}
+\frac {8193909687039684608000 } {81}\frac {\partial^{9}} {\partial z^{9}}\cr\cr
& + &\frac {470461823260549120000 } {243}\frac {\partial^{10}} {\partial z^{10}}
+\frac {2038677140211712000 } {81}\frac {\partial^{11}} {\partial z^{11}}\cr\cr
& + &\frac {48506933995520000 } {243}\frac {\partial^{12}} {\partial z^{12}}
+\frac {533043230720000 } {729}\frac {\partial^{13}} {\partial z^{13}}
\cr\cr
{\Lambda}_{30}&=& -\frac{2159303151863150104229789553754112}{31}
-48654707914751800750349947863040 \frac {\partial} {\partial z}\cr\cr
&-&\frac{511880813311501370486885415485440}{33}\frac {\partial^{2}} {\partial z^{2}}
-\frac{8986635138885396265944314675200}{3}\frac {\partial^{3}} {\partial z^{3}}\cr\cr
&-&\frac{8231837030002592217813488107520}{21}\frac {\partial^{4}} {\partial z^{4}}
-\frac{994048806800016269713608704000}{27}\frac {\partial^{5}} {\partial z^{5}}\cr\cr
&-&\frac{23064401700205672423804928000}{9}\frac {\partial^{6}} {\partial z^{6}}
-\frac{1210300901817271233249280000}{9}\frac {\partial^{7}} {\partial z^{7}}\cr\cr
&-&\frac{144529270268724971069440000}{27}\frac {\partial^{8}} {\partial z^{8}}
-\frac{13046432110823543111680000}{81}\frac {\partial^{9}} {\partial z^{9}}\cr\cr
&-&\frac{292528279332012179456000}{81}\frac {\partial^{10}} {\partial z^{10}}
-\frac{14240637964156764160000}{243}\frac {\partial^{11}} {\partial z^{11}}\cr\cr
&-&\frac{158354350809374720000}{243}\frac {\partial^{12}} {\partial z^{12}}
-\frac{1082077758361600000}{243}\frac {\partial^{13}} {\partial z^{13}}-
\frac{30916507381760000}{2187}\frac {\partial^{14}} {\partial z^{14}}
\cr\cr
{\Lambda} _ {32} & = & \frac {2670076846682445889021020926403215360} {33}
+\frac {626083852835613233107496318329487360 } {11}\frac {\partial} {\partial z}\cr\cr
& + &\frac {1415511296960735013459232888430919680 } {77}\frac{\partial^{2}} {\partial z^{2}}\cr\cr
& + &\frac {7527223418779506297125766215730135040 } {2079}\frac{\partial^{3}} {\partial z^{3}}\cr\cr
& + &\frac {30670811902988610868733044395212800 } {63}\frac{\partial^{4}} {\partial z^{4}}
+\frac {2985869165188242712391948769034240 } {63}\frac {\partial^{5}}{\partial z^{5}}\cr\cr
& + &\frac {279845133841516681643554865152000} {81}\frac {\partial^{6}} {\partial z^{6}}
+\frac {1730210223516695470726140723200 } {9}\frac {\partial^{7}} {\partial z^{7}}\cr\cr
& + &\frac {74183184142509865313386496000 } {9}\frac {\partial^{8}} {\partial z^{8}}
+\frac {66248118236803414927228928000 } {243}\frac {\partial^{9}} {\partial z^{9}}\cr\cr
& + &\frac {559535835716648787476480000 } {81}\frac {\partial^{10}} {\partial z^{10}}
+\frac {10679832674656911933440000 } {81}\frac {\partial^{11}} {\partial z^{11}}\cr\cr
& + &\frac {4018789059467584962560000 } {2187}\frac {\partial^{12}} {\partial z^{12}}
+\frac {12888720929367162880000 } {729}\frac {\partial^{13}} {\partial z^{13}}\cr\cr
& + &\frac {76672938306764800000 } {729}\frac {\partial^{14}} {\partial z^{14}}
+\frac {1916823457669120000 } {6561}\frac {\partial^{15}} {\partial z^{15}}
\cr\cr
{\Lambda}_{34}&=&- \frac{3741636967673446400392834119273566830592}{35}\cr\cr
&-&\frac{4767193737067336747135706468201325264896}{63}\frac{\partial} {\partial z}\cr\cr
&-&\frac{519155507401854532031062088426877485056}{21}\frac{\partial^{2}} {\partial z^{2}}\cr\cr
&-&\frac{312071961106560483063539014737326243840}{63}\frac {\partial^{3}} {\partial z^{3}}\cr\cr
&-&\frac{128903404125961567159485765048447139840}{189}\frac {\partial^{4}} {\partial z^{4}}\cr\cr
&-&\frac{4314909366918936150528150700950814720}{63}\frac {\partial^{5}} {\partial z^{5}}\cr\cr
&-&\frac{981744920544500723759478554507345920}{189}\frac {\partial^{6}} {\partial z^{6}}
-\frac{24607702301966583754762684923904000 }{81}\frac {\partial^{7}} {\partial z^{7}}\cr\cr
&-&\frac{124753205800785643189073625088000}{9}\frac {\partial^{8}} {\partial z^{8}}
-\frac{40140655844966041077573386240000}{81}\frac {\partial^{9}} {\partial z^{9}}\cr\cr
&-&\frac{10097202925543228306914875801600}{729}\frac {\partial^{10}} {\partial z^{10}}
-\frac{72849472123630686490918912000}{243}\frac {\partial^{11}} {\partial z^{11}}\cr\cr
&-&\frac{3595327698638837279326208000}{729}\frac {\partial^{12}} {\partial z^{12}}
-\frac{130534558042367793233920000}{2187}\frac {\partial^{13}} {\partial z^{13}}\cr\cr
&-&\frac{365819732593856020480000}{729}\frac {\partial^{14}} {\partial z^{14}}
-\frac{5735135785346007040000}{2187}\frac {\partial^{15}} {\partial z^{15}}\cr\cr
&-&\frac{42170116068720640000}{6561} \frac {\partial^{16}} {\partial z^{16}}
\cr\cr
{\Lambda}_{36}&=&159413579338683186365024026439178514595840\cr\cr
&+&\frac{2155794756107046193412927737541100643287040}{19} \frac {\partial} {\partial z}\cr\cr
&+&\frac{112341575907547366000563570204224074547200}{3} \frac {\partial^{2}} {\partial z^{2}}\cr\cr
&+&\frac{22855104633955473663874140445533589995520}{3} \frac {\partial^{3}} {\partial z^{3}}\cr\cr
&+&\frac{3213037622910504190580403495825460428800}{3} \frac {\partial^{4}} {\partial z^{4}}\cr\cr
&+&\frac{6962335067186410948979088074137721896960}{63} \frac {\partial^{5}} {\partial z^{5}}\cr\cr
&+&\frac{78073382527872228413161946715337523200}{9} \frac {\partial^{6}} {\partial z^{6}}\cr\cr
&+&\frac{4766188157241647377684667181812940800}{9} \frac {\partial^{7}} {\partial z^{7}}\cr\cr
&+&\frac{2063692531136353540681569606270976000}{81} \frac {\partial^{8}} {\partial z^{8}}\cr\cr
&+&\frac{78760717728549843972252500000768000}{81} \frac {\partial^{9}} {\partial z^{9}}
+\frac{2386776658958665484957974528000000}{81} \frac {\partial^{10}} {\partial z^{10}}\cr\cr
&+&\frac{171480228039448377254453248000000}{243} \frac {\partial^{11}} {\partial z^{11}}
+\frac{3208895131106247468261376000000}{243} \frac {\partial^{12}} {\partial z^{12}}\cr\cr
&+&\frac{45983601574243125952839680000}{243} \frac {\partial^{13}} {\partial z^{13}}
+\frac{1463549301062447536537600000}{729} \frac {\partial^{14}} {\partial z^{14}}\cr\cr
&+&\frac{10845141770089299312640000}{729} \frac {\partial^{15}} {\partial z^{15}}
+\frac{50182438121777561600000}{729} \frac {\partial^{16}} {\partial z^{16}}\cr\cr
&+&\frac{2951908124810444800000}{19683} \frac {\partial^{17}} {\partial z^{17}}
\cr\cr
{\Lambda}_{38}&=&-\frac{10391445661043092211419204709119815967319785472}{39}\cr\cr
&-&190584672822232305490739133613865643302977536 \frac {\partial} {\partial z}\cr\cr
&-& 63473188850861150926529639847522509797195776 \frac {\partial^{2}} {\partial z^{2}}\cr\cr
&-&\frac{274873155197293443241431573476958710351790080}{21}\frac {\partial^{3}} {\partial z^{3}}\cr\cr
&-&\frac{13120852454297768354467583973540049594613760}{7}\frac {\partial^{4}} {\partial z^{4}}\cr\cr
&-&\frac{1386712176157067971154944352979352748032000}{7}\frac {\partial^{5}} {\partial z^{5}}\cr\cr
&-&\frac{3029027524095926996073667975259190290022400}{189}\frac {\partial^{6}} {\partial z^{6}}\cr\cr
&-&\frac{9140510746882305995676527482625589248000}{9}\frac {\partial^{7}} {\partial z^{7}}\cr\cr
&-&\frac{460328198681340945997263270337642496000}{9}\frac {\partial^{8}} {\partial z^{8}}\cr\cr
&-&\frac{501564150760484910313291868702310400000}{243}\frac {\partial^{9}} {\partial z^{9}}\cr\cr
&-&\frac{1807861877410055071439556490362880000}{27}\frac {\partial^{10}} {\partial z^{10}}\cr\cr
&-&\frac{15690883326747521747053191888896000}{9}\frac {\partial^{11}} {\partial z^{11}}
-\frac{8795874652295397476633678446592000}{243}\frac {\partial^{12}} {\partial z^{12}}\cr\cr
&-&591656087787052984688312320000 \frac {\partial^{13}} {\partial z^{13}}
-\frac{603718501700062364275179520000}{81}\frac {\partial^{14}} {\partial z^{14}}\cr\cr
&-&\frac{458576983990354153032908800000}{6561}\frac {\partial^{15}} {\partial z^{15}}
-\frac{1005934214046011615805440000}{2187}\frac {\partial^{16}} {\partial z^{16}}\cr\cr
&-&\frac{4150382823483485388800000}{2187}\frac {\partial^{17}} {\partial z^{17}}
-\frac{218441201235972915200000}{59049}\frac {\partial^{18}} {\partial z^{18}}
\cr\cr
{\Lambda} _ {40} & = & 496112904390319879115644797765931390770950963200\cr\cr
&+&\frac {1069331656090577937629556891318585566243297689600 } {3}\frac {\partial} {\partial z}\cr\cr
&+& 119682353491740411875623771856664419118612480000\frac {\partial^{2}} {\partial z^{2}}\cr\cr
&+&\frac {274797778772741716394855730070985043415531520000 } {11}\frac {\partial^{3}} {\partial z^{3}}\cr\cr
&+&\frac {32732071206704825695531333310095981904134144000 }{9}\frac {\partial^{4}} {\partial z^{4}}\cr\cr
&+&\frac {11779557297530084078798065293924074491215872006} {3}\frac {\partial^{5}} {\partial z^{5}}\cr\cr
&+&\frac {293631191599569071749238152406562312814592000 } {9}\frac {\partial^{6}} {\partial z^{6}}\cr\cr
&+&\frac {403808233363403033677574667447518270193664000 } {189}\frac {\partial^{7}} {\partial z^{7}}\cr\cr
&+&\frac {1007865530043686746608877524067599646720000 } {9}\frac {\partial^{8}} {\partial z^{1}}\cr\cr
&+&\frac {384161427306806462941628740665572773068800 } {81}\frac{\partial^{9}} {\partial z^{9}}\cr\cr
&+&\frac {39631116865603436721213013206220144640000 } {243}\frac {\partial^{10}} {\partial z^{10}}\cr\cr
&+&\frac {123042011557644139760550010539868160000} {27}\frac{\partial^{11}} {\partial z^{11}}\cr\cr
&+&\frac {8352866246784718998029139037388800000 } {81}\frac {\partial^{12}} {\partial z^{12}}\cr\cr
&+&\frac {4100086058603581260066617334169600000 } {2187}\frac {\partial^{13}} {\partial z^{13}}\cr\cr
&+&\frac {19686984349036652753017398886400000} {729}\frac{\partial^{14}} {\partial z^{14}} \cr\cr
&+&\frac {659302285447370301782180495360000 } {2187}\frac {\partial^{15}} {\partial z^{15}} \cr\cr
&+&\frac {16514678011322398435105177600000 } {6561}\frac {\partial^{16}} {\partial z^{16}}
+\frac {32388401730801267549798400000} {2187}\frac {\partial^{17}}{\partial z^{17}} \cr\cr
& + &\frac {1079099534105706201088000000} {19683}\frac {\partial^{18}} {\partial z^{18}}
+\frac {5679471232135295795200000 } {59049}\frac {\partial^{19}} {\partial z^{19}}\cr\cr
{\Lambda}_{42}&=&-\frac{44005099160990369880384937311469520877799732781514752}{43}\cr\cr
&-&\frac{8120604311924082466257444539016879959072375255859200}{11}\frac{\partial}{\partial z}\cr\cr
&-&\frac{24720905474527566655311425549260362875836889238077440}{99}\frac{\partial^{2}}{\partial z^{2}}\cr\cr
&-&\frac{1738922931467714284426916859056763275358516438630400}{33}\frac{\partial^{3}}{\partial z^{3}}\cr\cr
&-&\frac{256935186161150640345584414324081276343178074521600}{33}\frac{\partial^{4}}{\partial z^{4}}\cr\cr
&-&\frac{7711045988353616624443279030923560745778020352000}{9}\frac{\partial^{5}}{\partial z^{5}}\cr\cr
&-&\frac{32151648261904850027188031660163591633590288384000}{441}\frac{\partial^{6}}{\partial z^{6}}\cr\cr
&-&\frac{44240307277911206936767084456248410215599308800}{9}\frac{\partial^{7}}{\partial z^{7}}\cr\cr
&-&\frac{7205847810440158345676060152056967530650009600}{27}\frac{\partial^{8}}{\partial z^{8}}\cr\cr
&-&\frac{955067662920105432280241864301397697363968000}{81}\frac{\partial^{9}}{\partial z^{9}}\cr\cr
&-&\frac{34546223084159754404936047659812729598771200}{81}\frac{\partial^{10}}{\partial z^{10}}\cr\cr
&-&\frac{27689949508193878147289960218372129423360000}{2187}\frac{\partial^{11}}{\partial z^{11}}\cr\cr
&-&\frac{224640158874807669136067409519161114624000}{729}\frac{\partial^{12}}{\partial z^{12}}\cr\cr
&-&\frac{4461077328174592618450418594527313920000}{729}\frac{\partial^{13}}{\partial z^{13}}\cr\cr
&-&\frac{214898355772427706883972917483274240000}{2187}\frac{\partial^{14}}{\partial z^{14}}\cr\cr
&-&\frac{2749287295891060396344910964326400000}{2187}\frac{\partial^{15}}{\partial z^{15}}\cr\cr
&-&\frac{27388396419147164765094982451200000}{2187}\frac{\partial^{16}}{\partial z^{16}}
-\frac{614793003547068961392223846400000}{6561}\frac{\partial^{17}}{\partial z^{17}}\cr\cr
&-&\frac{9760023646172470306360524800000}{19683}\frac{\partial^{18}}{\partial z^{18}}
-\frac{32600164872456597864448000000}{19683}\frac{\partial^{19}}{\partial z^{19}}\cr\cr
&-&\frac{465716641035094255206400000}{177147}\frac{\partial^{20}}{\partial z^{20}}\cr\cr
{\Lambda}_{44}&=&\frac{20943579517376089565211108276446305156743047301189074944}{9}\cr\cr
&+&\frac{116254688573101425742971225840997661055052901418740482048}{69}\frac{\partial}{\partial z}\cr\cr
&+&\frac{1721139709109764951295165469614030190500262281058713600}{3}\frac{\partial^{2}}{\partial z^{2}}\cr\cr
&+&\frac{3301381843756425815293186804565761697466537358340915200}{27}\frac{\partial^{3}}{\partial z^{3}}\cr\cr
&+&\frac{384509335313339043400399860022570820818290930614272000}{21}\frac{\partial^{4}}{\partial z^{4}}\cr\cr
&+&\frac{43045789687467854164482525203549027527256447119261696}{21}\frac{\partial^{5}}{\partial z^{5}}\cr\cr
&+&\frac{33678248861268060666775025587247310004252240919920640}{189}\frac{\partial^{6}}{\partial z^{6}}\cr\cr
&+&\frac{5438124340157395665299529611131812425934383328788480}{441}\frac{\partial^{7}}{\partial z^{7}}\cr\cr
&+&\frac{6217296215843337958044848270506292627122002329600}{9}\frac{\partial^{8}}{\partial z^{8}}\cr\cr
&+&\frac{69288197294497084080368221343348306002732226969600}{2187}\frac{\partial^{9}}{\partial z^{9}}\cr\cr
&+&\frac{873315012104334881596233231941443505339603353600}{729}\frac{\partial^{10}}{\partial z^{10}}\cr\cr
&+&\frac{27326721652816619929511408200507898280489779200}{729}\frac{\partial^{11}}{\partial z^{11}}\cr\cr
&+&\frac{6372332647838686488084841827292349152624640000}{6561}\frac{\partial^{12}}{\partial z^{12}}\cr\cr
&+&\frac{15153453612897392631539944484776121466880000}{729}\frac{\partial^{13}}{\partial z^{13}}\cr\cr
&+&\frac{266328927157220380709911018551220633600000}{729}\frac{\partial^{14}}{\partial z^{14}}\cr\cr
&+&\frac{34252430009882232551799885015875059712000}{6561}\frac{\partial^{15}}{\partial z^{15}}\cr\cr
&+&\frac{130619085807916829446230651754250240000}{2187}\frac{\partial^{16}}{\partial z^{16}}\cr\cr
&+&\frac{1168507119768076415864656843243520000}{2187}\frac{\partial^{17}}{\partial z^{17}}\cr\cr
&+&\frac{638323031325250579614695791001600000}{177147}\frac{\partial^{18}}{\partial z^{18}}
+\frac{1018152514930769271297762918400000}{59049}\frac{\partial^{19}}{\partial z^{19}}\cr\cr
&+&\frac{3083975596934394157976780800000}{59049}\frac{\partial^{20}}{\partial z^{20}}
+\frac{40051631129018105947750400000}{531441}\frac{\partial^{21}}{\partial z^{21}}\cr\cr
&\vdots&
\nonumber\\
\label{L2pSupp}
\end{eqnarray}

\section{Kronecker functions ${\rm K}_{2p+2}^{\frac{1}{2},\frac{1}{2}}(\tau)$ for $p$ from $1$ up to $22$}
\label{KroneckerSupp}
The Kronecker function is defined in Eq.\ (13) and discussed in Appendix C of the paper.


\section{Expressions for $f_p(\rho)$ up to $p=22$ for $\rho =1$ and $\rho =2$.}
\label{expressionfp}

Here,
\begin{equation}
x = \frac{\pi^4}{16 \Gamma[3/4]^8} =  \frac{\Gamma[1/4]^8} {256 \pi^4}=1.1973169873731537\ldots
\end{equation}

For $\rho=1$ (square systems) we obtain for $f_p(1)=f_p^\mathrm{sq}$:
\begin{eqnarray}
f_0 &=& \ln 2\cr\cr
f_1 &=& \frac{\pi x}{20}\cr\cr
f_2 &=&\frac{\pi^2 x^2}{2\cdot3\cdot7}\cr\cr
f_3 &=& \pi^3 \left( \frac{x^3}{2\cdot3^2} + \frac{3 x^2}{2^4\cdot5^2} \right)\cr\cr
f_4 &=& \pi^4 \left( \frac{5 x^4}{3^3} + \frac{3\cdot17 x^3}{2\cdot5\cdot7\cdot11} \right)\ldots \cr\cr
f_5 &=&  \pi^5 \left( \frac{11\cdot173 x^5}{2^2\cdot3^4\cdot5} + \frac{19\cdot 31 x^4}{2^2\cdot5\cdot7^2} + \frac{3^2\cdot 7^3 x^3}{2^6\cdot5^3\cdot13} \right)\cr\cr
f_6 &=& \pi^6 \left( \frac{7\cdot1801x^6}{3^5 \cdot5} + \frac{13\cdot 37 x^5}{3^2\cdot7} + \frac{3^2\cdot 599 x^4}{2^4\cdot5\cdot7\cdot11} \right)\cr\cr
f_7 &=& \pi^7 \left( \frac{23\cdot43\cdot6841x^7}{2\cdot3^6 \cdot5\cdot 7} +\frac{127\cdot2411x^6}{2^3\cdot3^2 \cdot5 \cdot7} + \frac{37\cdot43\cdot67\cdot97 x^5}{2^5\cdot5^2\cdot7^2 \cdot11} + \frac{7^2\cdot9 \cdot11\cdot31\cdot127 x^4}{2^8\cdot5^4\cdot13\cdot17} \right)\cr\cr
f_8 &=& \pi^8\left(\frac{2^4\cdot17\cdot71\cdot251x^8}{3^7 } +\frac{197\cdot257\cdot467x^7}{2\cdot3^3 \cdot5^2\cdot 7} +\frac{17\cdot47\cdot89\cdot1117x^6}{2\cdot3 \cdot5 \cdot7^3\cdot11} + \frac{3^2\cdot7\cdot29\cdot191\cdot257 x^5}{2^5\cdot5^2\cdot11 \cdot13\cdot19}  \right) \cr\cr
&=& 1.931460218138871039005168\times 10^8\cr\cr
f_9 &=& \pi^9\left(\frac{19\cdot 331617493 x^9}{2^2\cdot 3^8\cdot 5}+\frac{73\cdot 10578509 x^8}{2\cdot 3^5\cdot 5^2}+\frac{13\cdot 17\cdot 19\cdot 59\cdot 281\cdot 293
	x^7}{2^6\cdot 5^2\cdot 7^2\cdot 11}\right.\cr\cr&+&\left.\
\frac{3\cdot 73\cdot 15629\cdot 54581 x^6}{2^3\cdot 5^3\cdot 7^2\cdot 11^2\cdot 13}+\frac{3^5\cdot 7^2\cdot 11\cdot 53\cdot 809 x^5}{2^8\cdot 5^5\cdot 13\cdot 17}\right)\cr\cr
f_{10} &=&\pi^{10}\left(\frac{11\cdot 19\cdot 31\cdot 163\cdot 1205669 x^{10}}{2\cdot 3^9\cdot 5^2}+\frac{41\cdot 571\cdot 47951 x^9}{3^4\cdot 7}
+\frac{31\cdot 41\cdot 131\cdot 1823033
	x^8}{2\cdot 3^3\cdot 5^3\cdot 7^2}\right.\cr\cr&+&\left.\frac{5\cdot 41\cdot 22447\cdot 71569 x^7}{2^4\cdot 3\cdot 7^3\cdot 11\cdot 13}
+\frac{3^2\cdot 7\cdot 31\cdot 47\cdot 163\cdot 389\cdot 11423 x^6}{2^9\cdot 5^2\cdot 11\cdot 13\cdot 17\cdot 19\cdot 23}\right)\cr\cr &=&2.081722986975406344810703\times 10^{12}
\cr\cr
f_{11} &=&\pi^{11}\left( \frac{2\cdot 71\cdot 683 \cdot7146270149 x^{11}}{3^{10}\cdot 5^2\cdot 11}+\frac{13\cdot 23\cdot 89\cdot 809\cdot 1021\cdot 2287 x^{10}}{2^4\cdot 3^5\cdot 5^2\cdot 7}\right.\cr\cr
&+&\left. \frac{683\cdot 1193\cdot 589881377
	x^9}{2^5\cdot 3^3\cdot 5^2\cdot 7^2\cdot  11}+\frac{23\cdot 47\cdot 89\cdot 7459\cdot 32488273 x^8}{2^6\cdot 5^3\cdot 7^4\cdot 11\cdot 13}+\frac{3\cdot 173\cdot 683\cdot 27127\cdot 696403
	x^7}{2^9\cdot 5^4\cdot 11^2\cdot 17\cdot 19}\right.\cr\cr
&+&\left.\frac{3^3\cdot 7^2\cdot 11^2\cdot 19\cdot 71\cdot 89\cdot 8647 x^6}{2^{12}\cdot 5^5\cdot 13^2\cdot 17}\right)\cr\cr
f_{12} &=&\frac{1494465815519638 \pi ^{12} x^{12}}{885735
	\text{}}+\frac{3750028264324378 \pi ^{12} x^{11}}{1148175
	\text{}}+\frac{31735297176739 \pi ^{12} x^{10}}{14850
	\text{}}\cr\cr
&+&\frac{1647525284191712077 \pi ^{12} x^9}{2942940000
	\text{}}+\frac{2294226615090622491 \pi ^{12} x^8}{42897500800
	\text{}}+\frac{11251112076439893 \pi ^{12} x^7}{9089600000
	\text{}}
\cr\cr
f_{13} &=&
\frac{95331073442028622739 \pi ^{13}
	x^{13}}{1209028275}+\frac{968115205530868763 \pi ^{13}
	x^{12}}{5740875}+\frac{33202587383195509369 \pi ^{13}
	x^{11}}{261954000}\cr\cr
&+&\frac{13433801250515300909 \pi ^{13}
	x^{10}}{331080750}+\frac{2080856407677473708977223 \pi ^{13}
	x^9}{390367257280000}\cr\cr
&+&\frac{4237033980286563458949 \pi ^{13}
	x^8}{18697307200000}+
\frac{71007637896117497421 \pi ^{13}
	x^7}{53322880000000}
\cr\cr
f_{14} &=&
\frac{109096008925578950455217 \pi ^{14}
	x^{14}}{25389593775}+\frac{34678739684277073619 \pi ^{14}
	x^{13}}{3444525}\cr\cr
&+&\frac{10080105263579353129921 \pi ^{14}
	x^{12}}{1178793000}+\frac{1229655220891603073131 \pi ^{14}
	x^{11}}{378378000}\cr\cr
&+&\frac{2259866999879603358576294563 \pi ^{14}
	x^{10}}{4098856201440000}+\frac{136765043511613720994919 \pi ^{14}
	x^9}{3788664880000}\cr\cr
&+&\frac{736268809289919291539937 \pi ^{14}
	x^8}{1245456992000000}\cr\cr
f_{15} &=&
\frac{29398426535990138452783109 \pi ^{15} x^{15}}{108812544750
	\text{}}+\frac{853288759158275254603801 \pi ^{15}
	x^{14}}{1240029000 \text{}}\cr\cr
&+&\frac{1398121841721926018979037 \pi
	^{15} x^{13}}{2143260000 \text{}}+\frac{373290939738048696968623
	\pi ^{15} x^{12}}{1297296000
	\text{}}\cr\cr
&+&\frac{6431031498310509293982161209 \pi ^{15}
	x^{11}}{106463797440000
	\text{}}+\frac{380464709619048264091161553041 \pi ^{15}
	x^{10}}{69064976288000000
	\text{}}\cr\cr
&+&\frac{192361993215691052073814839 \pi ^{15}
	x^9}{1143835264000000 \text{}}+\frac{26287947801806586870759939
	\pi ^{15} x^8}{36259558400000000 \text{}}
\cr\cr
f_{16} &=&
\frac{6970032002817348987779910304 \pi ^{16} x^{16}}{359081397675
	\text{}}+\frac{1506595884499780218427798961 \pi ^{16}
	x^{15}}{28210659750 \text{}}\cr\cr
&+&\frac{14830787156588356712006837 \pi
	^{16} x^{14}}{265228425
	\text{}}+\frac{60367468829971823095779228127 \pi ^{16}
	x^{13}}{2145403260000 \text{}}\cr\cr
&+&\frac{9802190335062947368446372827
	\pi ^{16} x^{12}}{1383917535000
	\text{}}+\frac{101978907091894334901035573751041 \pi ^{16}
	x^{11}}{120863708504000000
	\text{}}\cr\cr
&+&\frac{22098424275209630295826237223 \pi ^{16}
	x^{10}}{550470720800000
	\text{}}+\frac{70581397469374089895674939939 \pi ^{16}
	x^9}{144473011072000000 \text{}}
\cr\cr
f_{17} &=&
\frac{578837946949360182132048740122783 \pi ^{17}
	x^{17}}{366263025628500}+\frac{197660525994270284121305971067 \pi
	^{17} x^{16}}{42315989625}\cr\cr
&+&\frac{974051318302561036845143920967
	\pi ^{17}
	x^{15}}{181870920000}+\frac{4065305065841626734497432102699 \pi
	^{17}
	x^{14}}{1340877037500}\cr\cr
&+&\frac{77633729124438258734285882484173 \pi
	^{17}
	x^{13}}{86867439120000}+\frac{164552649811683260137503161848729307
	9 \pi ^{17}
	x^{12}}{12373422158097000000}\cr\cr
&+&\frac{411808486286195999835843617009
	7631 \pi ^{17}
	x^{11}}{462395405472000000}+\frac{29648409441993847728464028898031
	1 \pi ^{17}
	x^{10}}{1451675928560000000}\cr\cr
&+&\frac{5861259201268701521898085500644
	7 \pi ^{17} x^9}{87204237952000000000}
\cr\cr
f_{18} &=&
\frac{11359989704116579159015658014016699 \pi ^{18}
	x^{18}}{78484934063250}\cr\cr
&+&\frac{147399591669521771231589890651794
	\pi ^{18}
	x^{17}}{322252536375}\cr\cr
&+&\frac{49144524536237535186165069823538 \pi
	^{18} x^{16}}{86199238125}\cr\cr
&+&\frac{32314215114769174161769239247 \pi
	^{18} x^{15}}{90016920}\cr\cr
&+&\frac{20655483584419761996508307967636809
	\pi ^{18}
	x^{14}}{169816046400000}\cr\cr
&+&\frac{616101590352240498846031893513611
	\pi ^{18}
	x^{13}}{28151484900000}\cr\cr
&+&\frac{285515037629068791338315379166504420
	33 \pi ^{18}
	x^{12}}{14781949612430000000}\cr\cr
&+&\frac{451024541553800750659872127873
	195371 \pi ^{18}
	x^{11}}{6471418481528000000}\cr\cr
&+&\frac{2470129475798990766624527431525
	527 \pi ^{18} x^{10}}{3778524904960000000}
\cr\cr
f_{19} &=&
\frac{463931422343866247870943796574546813423 \pi ^{19}
	x^{19}}{31315488691236750}\cr\cr
&+&\frac{556864073653031507574302655876149
	2753 \pi ^{19}
	x^{18}}{111714212610000}\cr\cr
&+&\frac{24996270506804428401358963645035273
	37 \pi ^{19}
	x^{17}}{37238070870000}\cr\cr
&+&\frac{134874441736003428518767637882592369
	79 \pi ^{19}
	x^{16}}{289629440100000}\cr\cr
&+&\frac{40147924140721988039982890427278583
	9421 \pi ^{19}
	x^{15}}{22466662938720000}\cr\cr
&+&\frac{367742669833522888733973725306481
	16194167 \pi ^{19}
	x^{14}}{9696722670835200000}\cr\cr
&+&\frac{2863091475729448191984541400856
	06736730888557 \pi ^{19}
	x^{13}}{679449357545422464000000}\cr\cr
&+&\frac{27956262763080395339054520
	40395273698206293 \pi ^{19}
	x^{12}}{129557798000190560000000}\cr\cr
&+&\frac{19747010876349868231661501
	901017548572207 \pi ^{19}
	x^{11}}{51331260833881600000000}\cr\cr
&+&\frac{712573113127820935847752998
	31861034703 \pi ^{19} x^{10}}{71507475120640000000000}
\cr\cr
f_{20} &=&
\frac{75728462084648409196111704401958993517 \pi ^{20}
	x^{20}}{44950462236225}\cr\cr
&+&\frac{251914279234463697541617633855905135
	173 \pi ^{20}
	x^{19}}{41892829728750}\cr\cr
&+&\frac{567052635952117514635943557644148376
	93 \pi ^{20}
	x^{18}}{6516662402250}\cr\cr
&+&\frac{8281465395691598216188747150802006845
	73 \pi ^{20}
	x^{17}}{125506090710000}\cr\cr
&+&\frac{28830030058122111263631604096693342
	147 \pi ^{20}
	x^{16}}{10176412806000}\cr\cr
&+&\frac{167369030521578970151687514419912609
	23916581 \pi ^{20}
	x^{15}}{24063558098580000000}\cr\cr
&+&\frac{179466852505861496779532268897
	63426140455373 \pi ^{20}
	x^{14}}{190754498776121280000}\cr\cr
&+&\frac{12407061379332839965609903011
	5834592497006011 \pi ^{20}
	x^{13}}{19399028577568640000000}\cr\cr
&+&\frac{324525808507762023228627309
	14063219081613 \pi ^{20}
	x^{12}}{179520321760197017600}\cr\cr
&+&\frac{42301569066432502821914339851
	64516712009 \pi ^{20} x^{11}}{3145405566318080000000}
\cr\cr
f_{21} &=&
\frac{31408406869364696247454369808245940218337171 \pi ^{21}
	x^{21}}{148336525379542500}\cr\cr
&+&\frac{62717596842153743424690888327603
	7627581529 \pi ^{21}
	x^{20}}{784849340632500}\cr\cr
&+&\frac{10911539185921582743218306451350889
	66989873 \pi ^{21}
	x^{19}}{882255832920000}\cr\cr
&+&\frac{97797104579952793733152498796222332
	12049 \pi ^{21}
	x^{18}}{9605057962500}\cr\cr
&+&\frac{5167169957462985091269570715652174604
	558916067 \pi ^{21}
	x^{17}}{10671664895892000000}\cr\cr
&+&\frac{155049809605025207724429367902
	7254387594749 \pi ^{21}
	x^{16}}{11452100660000000}\cr\cr
&+&\frac{121698936617485293799804280991595
	3054288650688171 \pi ^{21}
	x^{15}}{55998573424073280000000}\cr\cr
&+&\frac{476426313497190544317511319
	38431162170100145840573 \pi ^{21}
	x^{14}}{25451040518055616464000000}\cr\cr
&+&\frac{342288364531070015862353
	749985968984581231938181 \pi ^{21}
	x^{13}}{4527842434928046080000000}\cr\cr
&+&\frac{4445156774251037838650884
	82993433086160830080623 \pi ^{21}
	x^{12}}{413001991712896102400000000}\cr\cr
&+&\frac{25728970528006901728793
	64946467847811060379 \pi ^{21} x^{11}}{1144119601930240000000000}
\cr\cr
f_{22} &=&
\frac{3256501953006311534555567816069085633410604452 \pi ^{22}
	x^{22}}{111252394034656875
	\text{}}\cr\cr
&+&\frac{223946357017483872773158590693529918532092388 \pi
	^{22} x^{21}}{1922880884549625
	\text{}}\cr\cr
&+&\frac{421143517975157644803403371373667089286071181387
	\pi ^{22} x^{20}}{2193508564597350000
	\text{}}\cr\cr
&+&\frac{2890459416990614686828985811506542900102829801 \pi
	^{22} x^{19}}{16943322245850000
	\text{}}\cr\cr
&+&\frac{22248401942730587330439556166694332034489158297353
	\pi ^{22} x^{18}}{249716958563872800000
	\text{}}\cr\cr
&+&\frac{729277418657979329101715426491332598353481941317
	\pi ^{22} x^{17}}{26032394670282000000
	\text{}}\cr\cr
&+&\frac{93116193270424975807889040435972509128387022437027
	\pi ^{22} x^{16}}{17803894267979820000000
	\text{}}\cr\cr
&+&\frac{109187199968828094459570591384580902750130885654532
	973 \pi ^{22} x^{15}}{197952537362654794720000000
	\text{}}\cr\cr
&+&\frac{195933036861285685806192115835766304797369062523349
	\pi ^{22} x^{14}}{6586215818069888000000000
	\text{}}\cr\cr
&+&\frac{102765542403746630299897870370973192033242550324477
	\pi ^{22} x^{13}}{152092472600357825536000000
	\text{}}\cr\cr
&+&\frac{374052897414734765464410254427521007628363410381
	\pi ^{22} x^{12}}{91693295614695424000000000 \text{}}\nonumber
\end{eqnarray}
For $\rho=2$ we obtain

\section{The coefficients $f_{p}^{\,\mathrm {strip}}$ for infinitely long strip up to 22-nd order.}
\begin{eqnarray}\cr\cr
f_0^{\,\mathrm{strip}}&=&\frac{\pi }{24}\cr\cr
f_1^{\,\mathrm {strip}}&=&\frac{7\pi^3}{2880}\cr\cr
f_2^{\,\mathrm {strip}}&=&\frac{31 \pi ^5}{48384}\cr\cr
f_3^{\,\mathrm {strip}}&=&\frac{10033 \pi ^7}{19353600}\cr\cr
f_4^{\,\mathrm {strip}}&=&\frac{35989 \pi ^9}{43794432}\cr\cr
f_5^{\,\mathrm {strip}}&=&\frac{180686706457 \pi ^{11}}{83691159552000}\cr\cr
f_6^{\,\mathrm {strip}}&=&\frac{1248808051 \pi ^{13}}{147149291520}\cr\cr
f_7^{\,\mathrm {strip}}&=&\frac{18201691489278203 \pi ^{15}}{390239921111040000}\cr\cr
f_8^{\,\mathrm {strip}}&=&\frac{10510145076561543761 \pi ^{17}}{30774779284606156800}\cr\cr
f_9^{\,\mathrm {strip}}&=&\frac{2641801043421964301500517 \pi ^{19}}{822126246603050188800000}\cr\cr
f_{10}^{\,\mathrm {strip}}&=&\frac{890644111024794636680297 \pi ^{21}}{23574737045448504115200}\cr\cr
f_{11}^{\,\mathrm {strip}}&=&\frac{3765413304447846570644621549257363 \pi ^{23}}{6937903664053222070078668800000}\cr\cr
f_{12}^{\,\mathrm {strip}}&=&\frac{12907952448064394680137004372229 \pi
   ^{25}}{1379922164796026467123200000}\cr\cr
f_{13}^{\,\mathrm {strip}}&=&\frac{16896084244127140258410059328873017845607 \pi
   ^{27}}{88691895003325271524434247680000000}\cr\cr
f_{14}^{\,\mathrm {strip}}&=&\frac{777085597428101473891367515258640943820437683 \pi
   ^{29}}{171702126045913385930688048188620800000}\cr\cr
f_{15}^{\,\mathrm {strip}}&=&\frac{1090962143004132505971333057353842612185864968769941 \pi
   ^{31}}{8794716689936134988449787818908057600000000}\cr\cr
f_{16}^{\,\mathrm {strip}}&=&\frac{3877665028080307480532041835292907036827716659661 \pi
   ^{33}}{998003484340148998237978028408832000000}\cr\cr
f_{17}^{\,\mathrm {strip}}&=&\frac{83262171237500203913591200756651666001878112953167800379583
   1821 \pi
   ^{35}}{6037141419686122281072868728221892882726912000000000}\cr\cr
f_{18}^{\,\mathrm {strip}}&=&\frac{6837494555647359213581019858108738068456017538362703473391
   \pi ^{37}}{1241580715009169630793819594729581445120000000}\cr\cr
f_{19}^{\,\mathrm {strip}}&=&\frac{40640212516033772542598227204848498226097217251235300469544
   3900075719429 \pi
   ^{39}}{1653654351450743032392093603708152510564668538880000000
   000}\cr\cr
f_{20}^{\,\mathrm {strip}}&=&\frac{21889914588055007238570026130471433324750152112684538372363
   7174989637757 \pi
   ^{41}}{1796404614763434476661701334109351746076428730368000000
   0}\cr\cr
f_{21}^{\,\mathrm {strip}}&=&\frac{46701889227647792948664808540770377137800777804895803823742
   6016207399125129834551 \pi
   ^{43}}{6993839434966207251731192564011704332899584180014284800
   00000000}\cr\cr
f_{22}^{\,\mathrm {strip}}&=&\frac{72398587772918392423264065285647729514516935483287691393635
   9798251863574799341509 \pi
   ^{45}}{1798676784606651273937514505439250527417310617849036800
   0000000}\cr\cr
\nonumber
\end{eqnarray}

\end{document}